\documentclass[times]{stvrauth}

\usepackage{tikz}
\usetikzlibrary{arrows,automata,positioning,trees,shapes,
decorations.pathreplacing,trees}
\usepackage{cite}
\usepackage{amsmath,amssymb,amsfonts}
\usepackage{algorithmic}
\usepackage{graphicx}
\usepackage{textcomp}
\usepackage[perpage]{footmisc}
\usepackage{algorithm}
\usepackage{array}
\usepackage{lineno,hyperref}
\usepackage{listings}
\usepackage{mathtools}
\usepackage{multirow}
\usepackage{refcount}
\usepackage{tabularx}
\usepackage{subfig}
\usepackage{caption}
\usepackage{float}
\modulolinenumbers[5]
\usepackage{color}
\usepackage{wrapfig}

\usepackage{txfonts}
\usepackage{booktabs}

\definecolor{keyword-col}{HTML}{033500}
\definecolor{comment-col}{HTML}{424541}
\definecolor{number-col}{HTML}{0343dF}
\definecolor{string-col}{HTML}{840000}

\lstdefinestyle{customc}{
  numbers=left,
  stepnumber=1,
  firstnumber=1,
  numberstyle=\tiny,
  framexleftmargin=1em,
  belowcaptionskip=1\baselineskip,
  breaklines=true,
  frame=tb,
  language=C++,
  showstringspaces=false,
  basicstyle=\tiny\ttfamily,
  keywordstyle=\color{keyword-col}\bfseries,
  commentstyle=\color{comment-col}\itshape,
  stringstyle=\color{string-col},
}

\usepackage[inline]{enumitem}
\setlist[itemize]{leftmargin=*}
\setlist[enumerate]{leftmargin=*}
\newlist{inline-list}{enumerate*}{1}
\setlist[inline-list]{itemjoin={{; }},itemjoin*={{, and }},label=\it(\roman*)}
\setlist[description]{leftmargin=0em}

\newcommand*{\MyDef}{\mathrm{def}}
\newcommand*{\eqdef}{\ensuremath{\mathop{\overset{\MyDef}{=}}}}

\newcommand\BibTeX{{\rmfamily B\kern-.05em \textsc{i\kern-.025em b}\kern-.08em
T\kern-.1667em\lower.7ex\hbox{E}\kern-.125emX}}

\def\volumeyear{2021}

\begin{document}

\title{Model Checking C++ Programs}
\author{Felipe R. Monteiro$^{1}$, Mikhail R. Gadelha$^{2}$, and Lucas C. Cordeiro$^{3}$}
\address{\centering $^{1}$Federal University of Amazonas, Manaus, Brazil \\
                                $^{2}$Igalia, A Coruña, Spain \\
                                $^{3}$University of Manchester, Manchester, United Kingdom}

\corraddr{felipemonteiro@ufam.edu.br}

\begin{abstract}
In the last three decades, memory safety issues in system programming languages such as C or C++ have been one of the significant sources of security vulnerabilities. However, there exist only a few attempts with limited success to cope with the complexity of C++ program verification. Here we describe and evaluate a novel verification approach based on bounded model checking (BMC) and satisfiability modulo theories (SMT) to verify C++ programs formally. Our verification approach analyzes bounded C++ programs by encoding into SMT various sophisticated features that the C++ programming language offers, such as templates, inheritance, polymorphism, exception handling, and the Standard C++ Libraries. We formalize these features within our formal verification framework using a decidable fragment of first-order logic and then show how state-of-the-art SMT solvers can efficiently handle that. We implemented our verification approach on top of ESBMC. We compare ESBMC to LLBMC and DIVINE, which are state-of-the-art verifiers to check C++ programs directly from the LLVM bitcode. Experimental results show that ESBMC can handle a wide range of C++ programs, presenting a higher number of correct verification results. At the same time, it reduces the verification time if compared to LLBMC and DIVINE tools. Additionally, ESBMC has been applied to a commercial C++ application in the telecommunication domain and successfully detected arithmetic overflow errors, potentially leading to security vulnerabilities.\\
\end{abstract}


\maketitle

%----------------------------------
\section{Introduction}
\label{intro}
%----------------------------------

Formal verification techniques can significantly positively impact software reliability as security becomes a significant concern~\cite{icse2020}. For more than 30 years now, memory safety issues in system programming languages such as C or C++ have been among the major sources of security vulnerabilities~\cite{SoK2013}. For instance, the Microsoft Security Response Center reported that approximately $70$\% of their security vulnerabilities every year are due to memory safety issues in their C and C++ code~\cite{MattMiller2019}. Beyond memory safety, undefined behavior (\textup{e.g.}, signed integer overflow) also represents another crucial source of errors that could potentially lead to security issues~\cite{OriginalSin2019}.

Software verification plays an essential role in ensuring overall product reliability. Over the last 15 years, formal techniques dramatically evolved~\cite{Clarke2018}, its adoption in industry has been growing~\cite{aws2018, Distefano2019, Sadowski2018}, and several tools to formally verify C programs have been proposed~\cite{svcomp2019}. However, there exist only a few attempts with limited success to cope with the complexity of C++ program verification~\cite{MerzFS12,Yang12,Blanc07,Prabhu11,divine4,STVR2017}. The main challenge here is to support sophisticated features that the C++ programming language offers, such as templates, sequential and associative template-based containers, strings \& streams, inheritance, polymorphism, and exception handling. Simultaneously, to be attractive for mainstream software development, C++ verifiers must handle large programs, maintain high speed and soundness, and support legacy designs.

In an attempt to cope with ever-growing system complexity, bounded model checking (BMC) based on satisfiability modulo theories (SMT) has been introduced as a complementary technique to Boolean satisfiability (SAT) for alleviating the state explosion problem~\cite{handbook09}. In this paper, we describe and evaluate a novel SMT-based BMC approach to verify C++ programs integrated into ESBMC~\cite{CordeiroFM12, MorseCN014, esbmc2018, esbmc2019}, a state-of-the-art context-bounded model checker. ESBMC can check for undefined behaviors and memory safety issues such as under- and overflow arithmetic, division-by-zero, pointer safety, array out-of-bounds violations, and user-defined assertions.

Our major contributions are twofold:
{\it (i)} we present a C++ operational model, an abstract representation of the Standard C++ Libraries that reflects their semantics and enables ESBMC to verify specific properties related to C++ structures (\textup{e.g.}, functional properties of standard containers) via function contracts (i.e., pre- and post-conditions), in addition to memory safety properties;
{\it (ii)} we also describe and evaluate novel approaches to handle exceptions in C++ programs (\textup{e.g.}, exception specification for functions and methods), which previous approaches could not handle~\cite{MerzFS12,Blanc07,Prabhu11}.
We also present an overview of ESBMC's type-checking engine and how it handles templates, inheritance, and polymorphism.
Finally, we compare our approach against LLBMC~\cite{MerzFS12}, a state-of-the-art bounded model checker based on SMT solvers, and DIVINE~\cite{divine4}, a state-of-the-art explicit-state model checker, both for C and C++ programs. Our experimental evaluation contains a broad set of benchmarks with over $1,500$ instances, where ESBMC reaches a success rate of $84.27$\%, outperforming LLBMC and DIVINE.

This article is a substantially revised and extended version of a previous contribution by Ramalho~\textup{et al.}~\cite{Ramalho13}. The major differences here are (\textit{i}) we extend the C++ operational model structure to handle new features from the Standard Template Libraries (STL) (\textup{e.g.}, associative template-based containers); (\textit{ii}) we provide details about the C++ rules used to throw and catch exceptions; \textit{(iii)} we support ${\tt terminate}$ and ${\tt unexpected}$ handlers; and \textit{(iv)} we extend approximately $36$\% our experimental evaluation with a completely new set of benchmarks.

The remainder of this article is organized as follows. Section~\ref{sec:background} gives a brief introduction to BMC and describes the background theories of the SMT solvers relevant to our contributions. In Section~\ref{sec:type-checking}, we describe the aspects of C++ handled in type-checking, i.e., our current approach to support templates and the mechanism to support inheritance and polymorphism features. We then present the main contributions, Section~\ref{sec:cpp-operational-model} presents the operational model to replace the STL in the verification process; and Section~\ref{sec:exception-handling} describes the exception handling encoding. Section~\ref{sec:experimental-evaluation} presents the results of our experimental evaluation, where we also compare our verification results to other state-of-the-art C++ model checkers. Finally, in Section~\ref{sec:related-work}, we discuss the related work, and we conclude in Section~\ref{sec:conclusion} along with our future research directions.

%-----------------------------------
\section{Background Theory}
\label{sec:background}
%-----------------------------------

ESBMC is a bounded model checker based on CProver framework~\cite{Clarke04} aimed to support SMT solvers natively. ESBMC generates verification conditions (VCs) for a given C or C++ program, encodes them using different SMT background theories (\textup{i.e.}, linear integer and real arithmetic and bit-vectors), and solvers (\textup{i.e.}, Boolector~\cite{NiemetzPreinerBiere-JSAT15}, Z3~\cite{Z08}, Yices~\cite{DBLP:conf/cav/2014}, MathSAT~\cite{Cimatti:2013:MSS:2450387.2450400}, and CVC4~\cite{Barrett:2011:CVC:2032305.2032319}). %, while CBMC relies mostly on bit-vectors and SAT solvers.
ESBMC represents one of the most prominent BMC tools for software verification, according to the last editions of the Intl. Competition on Software Verification (SV-COMP)~\cite{svcomp2020} and the Intl. Competition on Software Testing~\cite{TESTCOMP20}; in particular, it was ranked at the top three verifiers in the overall ranking of SV-COMP 2020~\cite{svcomp2020}. ESBMC has been applied to verify (embedded) software in digital filters~\cite{AbreuGCFS16} and digital controllers~\cite{BessaICF14}, and Unmanned Aerial Vehicles~\cite{ChavesBIFCF18}.

%----------------------------------
\subsection{Bounded Model Checking}
\label{sec:bmc}
%----------------------------------

In BMC, the program to be analyzed is modeled as a state transition system, which is extracted from the control-flow graph (CFG)~\cite{CFG}. This graph is built as part of a translation process from program code to static single assignment (SSA) form. A node in the CFG represents either a (non-) deterministic assignment or a conditional statement, while an edge in the CFG represents a possible change in the program's control location.

Given a transition system \textit{M}, a property $\phi$, and a bound \textit{k}, BMC unrolls the system \textit{k} times and translates it into a VC $\psi$, such that $\psi$ is satisfiable if and only if $\phi$ has a counterexample of length \textit{k} or less~\cite{handbook09}. The associated model checking problem is formulated by constructing the following logical formula:
\begin{equation}
\small
\label{bmceq}
  \psi_k = I(s_0) \wedge \bigwedge^{k-1}_{i=0} T(s_i, s_{i+1}) \wedge \bigvee^{k}_{i=0} \neg \phi(s_i),
\end{equation}

\noindent given that $\phi$ is a safety property, $I$ is the set of initial
states of $M$ and $T(s_i, s_{i+1})$ is the transition relation of $M$ between
steps $i$ and $i + 1$. Hence, $I(s_0) \wedge \bigwedge^{j-1}_{i=0} T(s_i,
s_{i+1})$ represents the executions of $M$ of length $j$ and the
formula~\eqref{bmceq} can be satisfied if and only if, for some $j \leq k$,
there exists a reachable state at step $j$ in which $\phi$ is violated. If the
formula~\eqref{bmceq} is satisfiable, then the SMT solver provides a satisfying
assignment, from which we can extract the values of the program variables to
construct a counterexample. A counterexample for a property $\phi$ is a
sequence of states $s_0,s_1, \cdots, s_k$ with $s_0 \in S_0$ and $T(s_i,
s_{i+1}) $ with $0 \leq i < k$.

If the formula~\eqref{bmceq} is unsatisfiable, we can conclude that no error
state is reachable in $k$ steps or less. In this case, BMC techniques are not
complete because there might still be a counterexample that is longer than
\textit{k}. Completeness can only be ensured if we know an upper bound on the depth of the state space. This means that if we can ensure that we have already explored all the relevant behavior of the system, and searching any deeper only exhibits states that have already been verified~\cite{Kroening2011}.

%-----------------------------------
\subsection{Satisfiability Modulo Theories}
\label{smt-theories}
%-----------------------------------

SMT decides the satisfiability of a fragment of quantifier-free first-order formulae using a combination of different background theories. It generalizes propositional satisfiability by supporting uninterpreted functions, linear and non-linear arithmetic, bit-vectors, tuples, arrays, and other decidable first-order theories. Given a theory $\tau$ and a quantifier-free formula $\psi$, we say that $\psi$ is $\tau$-satisfiable if and only if there exists a structure that satisfies both the formula and the sentences of $\tau$, or equivalently if ${\tau}\cup \{\psi\}$ is satisfiable~\cite{Bradley07}. Given a set $\Gamma\cup \{\psi\}$ of formulae over $\tau$, we say that $\psi$ is a $\tau$-consequence of $\Gamma$, and write $\Gamma\models_{\tau}\psi$, if and only if every model of ${\tau}\cup\Gamma$ is also a model of $\psi$.  Checking $\Gamma\models_{\tau}\psi$ can be reduced in the usual way to checking the $\tau$-satisfiability of $\Gamma\cup\{\neg\psi\}$.

ESBMC heavily uses the (non-extensional) theory of arrays $\mathcal{T}_\mathcal{A}$ based on the McCarthy axioms~\cite{McCarthy62}, to properly encode properties and behaviors of  the STL models ({\it cf.} Section~\ref{sec:cpp-operational-model}) and the C++ exception handling features ({\it cf.} Section~\ref{sec:exception-handling}). We define conditional expressions~\cite{boolector2015} over bitvectors using the $ite(c, t_1, t_2)$ operator, where $c$ is the condition expression, $t_1$ is the consequent branch $ite(\top,t_1,t_2) = t1$, and $t_2$ is the alternative branch $ite(\bot,t_1,t_2) =t_2$. The operation $select(a, i)$ denotes the value of an array $a$ at index position $i$ and $store(a, i, v)$ denotes an array that is exactly the same as array $a$ except that the value at index position $i$ is $v$. Formally, the functions $select$ and $store$ can then be characterized by the following two axioms~\cite{Barrett:2011:CVC:2032305.2032319,Brummayer:2009:BES:1532891.1532912,Z08}:
\begin{equation*}
\small
\begin{array}{r c l}
  i=j~    & \Rightarrow &\textit{select}(\textit{store}(a,i,v),j)=v \\
  \neg \: (i = j)& \Rightarrow &\textit{select}(\textit{store}(a,i,v),j)=\textit{select}(a,j)
\end{array}
\end{equation*}

Finally, an important component of our models is the {\it memcpy pattern} through lambda terms introduced by Preiner, Niemetz, and Biere~\cite{boolector2015}. It allows us to reason about operations over multiple indices without the need for quantifiers. Here, the $memcpy(a, b, i, k, n)$ operation denotes a copy of $n$ elements from array $a$ starting at position $i$ to array $b$ at the position $k$.

%-----------------------------------------------------------
\section{Static Type Checking of C++ Programs}
\label{sec:type-checking}
%-----------------------------------------------------------

\begin{figure*}[t]
\centering
\includegraphics[width=0.8\textwidth]{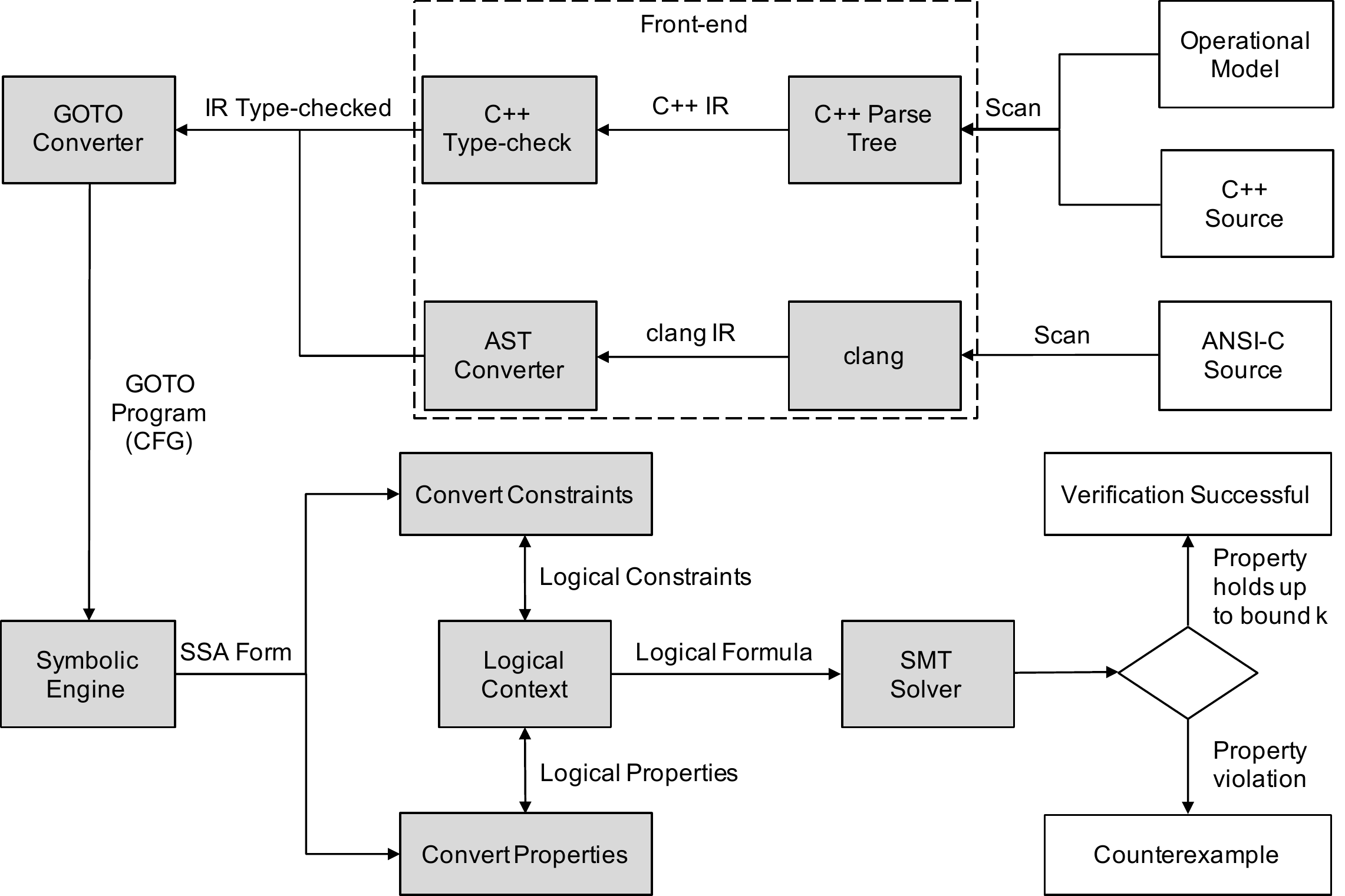}
\caption{ESBMC architectural overview. White rectangles
represent input and output; gray rectangles represent the steps of the
verification.
}
\label{figure:esbmc-arch}
\end{figure*}

The first steps when verifying C++ programs are the source-code parser and the type-checker, which are language-specific in ESBMC (see Fig.~\ref{figure:esbmc-arch}). For C++, the parser is heavily based on the GNU C++ Compiler (GCC)~\cite{GCC}, which allows ESBMC to find and report most of the syntax errors already reported by GCC. Type-checking provides all information used by the model; thus, a better type-checker means it is possible to model more programs. The code is statically analyzed on type-checking, including assignment checks, type-cast checks, pointer initialization checks, and function call checks. Furthermore, ESBMC handles three major C++ features on type-checking: template instantiation (i.e., after type-checking, all referenced templates are instantiated with concrete types), compile-time and runtime polymorphism, and inheritance (i.e., it replicates the methods and attributes of the base classes to the inherited class, which will have direct access).

By the end of the type-check, the Intermediate Representation (IR) creation is completed and used by the GOTO converter to generate the GOTO program. The verification of C programs is slightly different as it uses clang as a front-end to parse and type-check the program, as described in our previous work ~\cite{esbmc2018, esbmc2019}; the output, however, it is the same: a type-checked IR.

The GOTO converter converts the type-checked IR into GOTO expressions; this conversion
simplifies the IR of the original program (\textup{e.g.}, replacing of
${\tt switch}$ and ${\tt while}$ by ${\tt if}$ and ${\tt goto}$ statements). The
symbolic engine converts the GOTO program into
SSA form~\cite{cytron89} by unrolling loops up to bound $k$.
Assertions are inserted into the resulting SSA
expressions to verify memory-safety properties (\textup{e.g.}, array out-of-bounds access,
arithmetic under- and overflow, memory leaks, double frees, division-by-zero, etc.).
Also, most of the exception handling is carried out in this step, such
as the search for valid ${\tt catch}$, assignment of a thrown object to a valid
${\tt catch}$ object, replacement of ${\tt throw}$ statements
by GOTO expressions and exception specs for function calls ({\it cf.} Section~\ref{sec:exception-handling}).

Finally, two sets of quantifier-free formulae are created based on the SSA
expressions: $\mathcal{C}$ for the constraints and $\mathcal{P}$ for the properties,
as previously described. The two sets of formulae will be used as input for an
SMT solver that will produce a counterexample if there exists a violation of a given
property, or an unsatisfiable answer if the property holds.

%------------------------------------------
\subsection{Template Instantiation}
\label{sec:templates}
%------------------------------------------

Templates are not runtime objects~\cite{Stroustrup}. When a C++ program is compiled, classes and functions are generated from templates. Those templates are removed from the final executable. ESBMC has a similar process in which templates are only used until the type-checking phase, where all templates are instantiated and the classes and functions are generated. Any instantiated functions and classes are no longer templates. Hence, at the end of the type-checking phase, all templates are completely discarded. In ESBMC, the entire verification process of C++ programs, which make use of templates, is essentially split into two steps: \textit{creation of templates} and \textit{template instantiation}. The {\it creation of templates} is straightforward. It happens during the parsing step when all generic data types of the generated C++ IR are properly marked as ${\tt generic}$ and each specialization is paired with its corresponding primary template. No instantiated function or class is created during parsing because ESBMC does not know which template types will be instantiated.

A template instantiation happens when a template is used, instantiated with data types (\textup{e.g.}, ${\tt int}$, ${\tt float}$, or ${\tt string}$). ESBMC performs an in-depth search in the C++ IR during the type-checking process to trigger all instantiations. When a template instantiation is found, ESBMC firstly identifies which type of template it is dealing with (\textup{i.e.}, either ${\tt class}$ or ${\tt function}$ template) and which template arguments are used. It then searches whether an IR of that type was already created, \textup{i.e.}, whether its arguments have been previously instantiated. If so, no new IR is created; this avoids duplicating the IR, thus reducing the memory requirements of ESBMC. If there is no IR of that type, a new IR is created, used in the instantiation process, and saved for future searches. To create a new IR, ESBMC must select the most specialized template for the set of template arguments; therefore, ESBMC performs another search in the IR to select the proper template definition. ESBMC then checks whether there is a (partial or explicit) template specialization, matching the set of data types in the instantiation. If ESBMC does not find any template specialization, which matches the template arguments, it will select the primary template definition. Once the most specialized template is selected, ESBMC performs a transformation to replace all generic types for the data types specified in the instantiation; this transformation is necessary because, as stated previously, at the end of the C++ type-checking phase, all templates are removed.

\begin{wrapfigure}[20]{l}{7cm}
\centering
\begin{minipage}{0.4\textwidth}
\begin{lstlisting}
#include<cassert>
using namespace std;

// template creation
template <typename T>
bool qCompare(const T a, const T b) {
  return (a > b) ? true : false;
}

template <typename T>
bool qCompare(T a, T b) {
  return (a > b) ? true : false;
}

// template specialization
template<>
bool qCompare(float a, float b) {
  return (b > a) ? true : false;
}

int main() {
  // template instantiation
  assert((qCompare(1.5f, 2.5f)));
  assert((qCompare<int>(1, 2) == false));
  return 0;
}
\end{lstlisting}
\end{minipage}
\caption{Function template example.}
\label{figure:template-example}
\end{wrapfigure}

In order to concretely demonstrate the instantiation process in ESBMC, Fig.~\ref{figure:template-example} illustrates an example of function templates usage, which is based on the example ${\tt spec29}$ extracted from the GCC test suite.\footnote{https://github.com/nds32/gcc/} The first step, the template creation, happens when the declaration of a template function (lines $5$--$19$) is parsed. At this point, the generic IR of the template is created with a generic type. The second step, template instantiation, happens when the template is used. In Fig.~\ref{figure:template-example}, the template is instantiated twice (lines $23$ and $24$). It is also possible to determine the type implicitly (line $23$) or explicitly (line $24$). In implicit instantiation, the data type is determined by the types of the used parameters. In contrast, in the explicit instantiation, the data type is determined by the value passed between the $<$ and $>$ symbols.

\begin{figure}[h!]
   \centering
    \subfloat[Generic IR generated from the ${\tt qCompare}$ function template with generic type in
Fig.~\ref{figure:template-example}.\label{figure:template}]{%
      \tikzset{
  basic/.style  = {draw, text width=2cm, rectangle},
  root/.style   = {basic, rounded corners=2pt, thin, align=center, fill=gray!30},
  level 2/.style = {basic, rounded corners=2pt, thin, align=center, text width=5.5em},
  level 3/.style = {basic, dashed, align=center, text width=4.5em, fill=gray!30},
  level 4/.style = {basic, rounded corners=6pt, thin, align=center, text width=5.5em}
}

\begin{tikzpicture}[
  level 1/.style={sibling distance=40mm},
  edge from parent/.style={->,draw},
  >=latex, node distance=.7cm]

\node[root] (c2) {{\tt qCompare}};

% The second level, relatively positioned nodes
\begin{scope}[every node/.style={level 2}]
\node [below of = c2, xshift=15pt] (c21) {arguments};
\end{scope}

\begin{scope}[every node/.style={level 3}]
\node [below of = c21, xshift=15pt] (c211) {generic {\tt a}};
\node [below of = c211] (c212) {generic {\tt b}};
\end{scope}

\begin{scope}[every node/.style={level 2}]
\node [below of = c212, xshift=-15pt] (c22) {return};
\end{scope}

\begin{scope}[every node/.style={level 3}]
\node [below of = c22, xshift=15pt] (c221) {{\tt bool r}};
\end{scope}

\begin{scope}[every node/.style={level 2}]
\node [below of = c221, xshift=-15pt] (c23) {specialization};
\end{scope}

\begin{scope}[every node/.style={level 3}]
\node [below of = c23, xshift=15pt] (c231) {arguments};
\end{scope}

\begin{scope}[every node/.style={level 4}]
\node [below of = c231, xshift=15pt] (c241) {{\tt float r}};
\node [below of = c241] (c242) {{\tt float r}};
\end{scope}

\begin{scope}[every node/.style={level 3}]
\node [below of = c242, xshift=-15pt] (c232) {return};
\end{scope}

\begin{scope}[every node/.style={level 4}]
\node [below of = c232, xshift=15pt] (c244) {{\tt bool r}};
\end{scope}

\foreach \value in {1,...,3}
  \draw[->] (c2.195) |- (c2\value.west);

\foreach \value in {1,...,2}
 \draw[->] (c21.195) |- (c21\value.west);

\foreach \value in {1}
 \draw[->] (c22.195) |- (c22\value.west);

\foreach \value in {1,...,2}
  \draw[->] (c23.195) |- (c23\value.west);
  
\end{tikzpicture}
    }
    \hspace*{.4cm}%
    \subfloat[Instantiated IRs generated from the template instantiation with types ${\tt float}$ and ${\tt int}$ in
Fig.~\ref{figure:template-example}.\label{figure:template-instantiated}]{%
      \tikzset{
  basic/.style  = {draw, text width=2cm, rectangle},
  root/.style   = {basic, rounded corners=2pt, thin, align=center, fill=gray!30},
  level 2/.style = {basic, rounded corners=2pt, thin, align=center, text width=5.5em},
  level 3/.style = {basic, dashed, align=center, text width=4.5em, fill=gray!30},
  level 4/.style = {basic, rounded corners=6pt, thin, align=center}
}

\begin{tikzpicture}[
  level 1/.style={sibling distance=40mm},
  edge from parent/.style={->,draw},
  >=latex, node distance=.7cm]

\node[root] (c2) {{\tt qCompare}};

% The second level, relatively positioned nodes
\begin{scope}[every node/.style={level 2}]
\node [below of = c2, xshift=15pt] (c21) {arguments};
\end{scope}

\begin{scope}[every node/.style={level 3}]
\node [below of = c21, xshift=15pt] (c211) {{\tt float a}};
\node [below of = c211] (c212) {{\tt float b}};
\end{scope}

\begin{scope}[every node/.style={level 2}]
\node [below of = c212, xshift=-15pt] (c22) {return};
\end{scope}

\begin{scope}[every node/.style={level 3}]
\node [below of = c22, xshift=15pt] (c221) {{\tt bool r}};
\end{scope}

%second
\begin{scope}[every node/.style={root}]
\node [below of = c221, xshift=-30pt] (c3) {{\tt qCompare}};
\end{scope}

\begin{scope}[every node/.style={level 2}]
\node [below of = c3, xshift=15pt] (c31) {arguments};
\end{scope}

\begin{scope}[every node/.style={level 3}]
\node [below of = c31, xshift=15pt] (c311) {{\tt int a}};
\node [below of = c311] (c312) {{\tt int b}};
\end{scope}

\begin{scope}[every node/.style={level 2}]
\node [below of = c312, xshift=-15pt] (c32) {return};
\end{scope}

\begin{scope}[every node/.style={level 3}]
\node [below of = c32, xshift=15pt] (c321) {{\tt bool r}};
\end{scope}

\foreach \value in {1,...,2}
  \draw[->] (c2.195) |- (c2\value.west);

\foreach \value in {1,...,2}
 \draw[->] (c21.195) |- (c21\value.west);

\foreach \value in {1}
 \draw[->] (c22.195) |- (c22\value.west);
 
 % second
 \foreach \value in {1,...,2}
  \draw[->] (c3.195) |- (c3\value.west);

\foreach \value in {1,...,2}
 \draw[->] (c31.195) |- (c31\value.west);

\foreach \value in {1}
 \draw[->] (c32.195) |- (c32\value.west);

\vspace{1cm}

\end{tikzpicture}
    }
    \caption{Example of IR creation.}
    \label{figure:irep-exemplo}
  \end{figure}
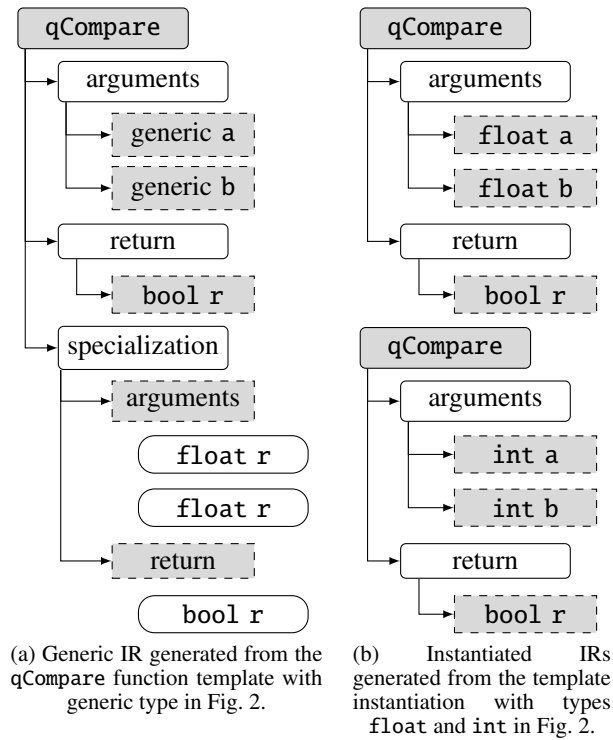

Fig.~\ref{figure:irep-exemplo} illustrates the generic IR and the instantiated IRs generated from the code in Fig.~\ref{figure:template-example}. Fig.~\ref{figure:template} illustrates the generic IR generated from the ${\tt qCompare}$ function template and its specialization, while Fig.~\ref{figure:template-instantiated} shows the IRs created from instantiating this template with data type ${\tt float}$ (line $23$) and ${\tt int}$ (line $24$). The function body is omitted in this figure, but it follows the same instantiation pattern. The generic IR is built with the function name, which is used as a key for future searches, the IR's arguments and return type, as can be seen in Fig.~\ref{figure:template}. Note that the data type is labeled as ${\tt generic}$, which means that the type is generic. In Fig.~\ref{figure:template-instantiated}, the data types that were previously labeled as ${\tt generic}$ are now labeled as ${\tt float}$ for the first instantiation and ${\tt int}$ for the second instantiation, which means that these instantiated IRs are not templates anymore and will not be removed at the end of the type-check phase. Finally, as described earlier, at the end of the type-check phase, the generic IR illustrated in Fig.~\ref{figure:template} is discarded.

After the template instantiation, the verification process resumes, as described by Cordeiro et al.~\cite{CordeiroFM10}. ESBMC is currently able to handle the verification of C++ programs
with template functions, class templates, and (partial and explicit) template specialization,
according to the C++03 standard~\cite{ISO14882:2003}.
The implementation of template instantiation in ESBMC is based
on the formalization previously presented by Siek and Taha~\cite{template2006} who introduced
the first proof of type safety of the template instantiation process for C++03 programs.

%------------------------------------------------
\subsection{Inheritance}
\label{sec:inheritance}
%------------------------------------------------

In contrast to Java, which only allows single inheritance~\cite{java2004}, where derived classes have only one base class, C++ also allows multiple inheritances, where a class may inherit from one or more unrelated base classes~\cite{Deitel}. This particular feature makes C++ programs harder to model check than programs in other object-oriented programming languages (\textup{e.g.}, Java) since it disallows the direct transfer of techniques developed for other, simpler programming languages~\cite{Pasareanu04,Anand:2007:JSE:1763507.1763523}. Multiple inheritance in C++ includes features that raise exciting challenges for model checking such as repeated and shared inheritance of base classes, object identity distinction, and dynamic dispatch~\cite{Ramananandro:2011}.

In ESBMC, inheritance is handled by
replicating the methods and attributes of the base classes to the derived class,
obeying the rules of inheritance defined in the C++03 standard~\cite{ISO14882:2003}. In particular, we follow these specifications to handle multiple inheritance and avoid issues such as name clashing when replicating the methods and attributes. For example, if two or more base classes implement a method that is not overridden by the derived class, every call to this method must specify which ``version'' inherited it is referring to. The rules are checked
in the type-check step of the verification (\textit{cf.}, Section~\ref{sec:type-checking}).

A formal description to represent the relationship between classes can be
described by a class hierarchy graph. This graph is represented by a
triple $\langle \mathit{C}, \prec_{s}, \prec_{r}\rangle$, where $\mathit{C}$ is
the set of classes, ${\prec_s} \subseteq {C \times C}$ refers to \textit{shared
inheritance} edges (i.e., if there exists a path from class $\mathit{X}$ to class $\mathit{Y}$ whose first edge is virtual),
and ${\prec_r} \subseteq {C \times C}$ are \textit{replicated inheritance} edges (i.e., if a class inherits from a base class that does not contain virtual
methods). We also define the set of all inheritance edges ${\prec_{sr}} = {\prec_s} \cup
{\prec_r}$. Thus, $\left( \mathit{C}, \leq_{sr} \right)$ is a partially
ordered set~\cite{Neggers99} and $\leq_{sr}$ is anti-symmetric (\textup{i.e.},
if one element $A$ of the set precedes $B$, the opposite relation cannot exist). {Importantly, during the replication process of all methods and attributes from the base classes to the derived ones, the inheritance model considers the access specifiers related to each component ({\it i.e.}, ${\tt public}$, ${\tt protected}$, and ${\tt private}$) and its friendship~\cite{Deitel}; therefore, we define two rules to deal with such restrictions: {\it (i)} only ${\tt public}$ and ${\tt protected}$ class members from base classes are joined in the derived class and {\it (ii)} if class $X \in C$ is a friend of class $Y \in C$, all private members in class $X$ are joined in class $Y$.}

\begin{wrapfigure}[15]{l}{8.5cm}
\centering
%\vspace{-12pt}
\includegraphics[scale=0.33,trim={0 2.3cm 0 2.8cm},clip]{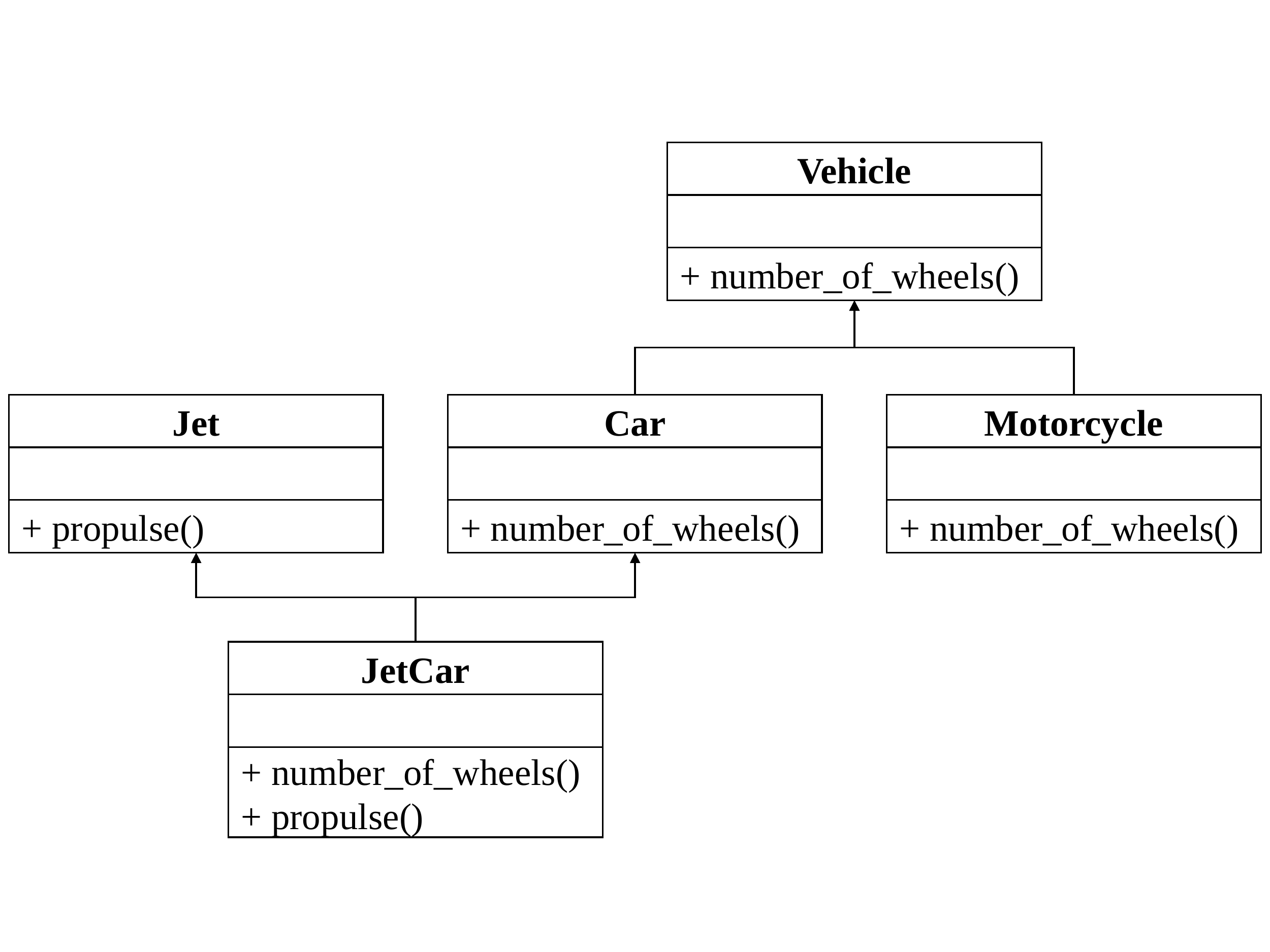}
\caption{${\tt Vehicle}$ class hierarchy UML diagram.}
\label{figure:uml_diagram}
\end{wrapfigure}

As an example, Fig.~\ref{figure:uml_diagram} shows an UML diagram that
represents the ${\tt Vehicle}$ class hierarchy, which contains multiple
inheritance. The replicated inheritance in the ${\tt JetCar}$ class relation
can be formalized by $\langle C, \emptyset, \{({\tt JetCar}, {\tt Car}),$
$({\tt JetCar}, {\tt Jet})\}\rangle$.

ESBMC creates an intermediate model for single and multiple inheritance,
handling replicated and shared inheritance where all classes are converted into structures and all methods and attributes of its parent classes are joined. This approach has the advantage of having direct access to the attributes and methods of the derived class and thus allows an easier validation, as the tool does not search for attributes or methods from base classes on each access. However, we replicate information to any new class, thus wasting memory resources.

In addition, we also support \textit{indirect inheritance}, where a class
inherits features from a derived class with one or more classes not directly connected. \textit{Indirect inheritance} is automatically handled due to our replication method: any derived class will already contain all methods and attributes from their base classes,
which will be replicated to any class that derives from them.
In Fig.~\ref{figure:uml_diagram}, we have
${\tt JetCar}\leq_{\mathit{sr}}{\tt Car}$ and
${\tt Car}\leq_{\mathit{sr}}{\tt Vehicle}$. Thus, the ${\tt JetCar}$
class can access features from the ${\tt Vehicle}$ class, but they are not
directly connected.

In object-oriented programming, the use of \textit{shared inheritance} is very
common~\cite{Deitel}. In contrast to other approaches (\textup{e.g.}, the one
proposed by Blanc, Groce, and Kroening~\cite{Blanc07}), ESBMC is able to verify
this kind of inheritance. A pure virtual class does not implement any method
and, if an object tries to create an instance of a pure virtual class, ESBMC
will fail with a ${\tt CONVERSION~ERROR}$ message (since it is statically checked during type-checking).

%------------------------------------------------
\subsection{Polymorphism}
\label{sec:polymorphism}
%------------------------------------------------

In order to handle polymorphism, \textup{i.e.}, allowing variable instances to
be bound to references of different types, related by
inheritance~\cite{Alexander02}, ESBMC implements a virtual function table
(i.e., ${\tt vtable}$) mechanism \cite{Driesen:1996:DCV:236337.236369}. When a class
defines a virtual method, ESBMC creates a ${\tt vtable}$, which contains a
pointer to each virtual method in the class. If a derived class
does not override a virtual method, then the pointers are copied to the
virtual table of the derived class. In contrast, if a derived class
overrides a virtual method, then the pointers in the virtual table of the derived class will point to the overridden method implementation.
Whenever a virtual method is called,
ESBMC executes the method pointed in the virtual table.
ESBMC also supports the unary scope resolution operator (i.e., ${\tt ::}$), which,
in this context, enables a derived class to access members from its parents,
a key component to support multiple inheritance.
\begin{figure}[!ht]
\centering
\begin{minipage}{0.5\textwidth}
\begin{lstlisting}[escapechar=^]
#include <cassert>

class Vehicle
{
public:
  Vehicle() {};
  virtual int number_of_wheels() = 0;
};

class Motorcycle : public Vehicle
{
public:
  Motorcycle() : Vehicle() {}; ^\label{motor_constr}^
  virtual int number_of_wheels() { return 2; };
};

class Car : public Vehicle
{
public:
  Car() : Vehicle() {}; ^\label{car_constr}^
  virtual int number_of_wheels() { return 4; };
};

int main()
{
  bool foo = nondet();

  Vehicle* v;
  if(foo)
    v = new Motorcycle();
  else
    v = new Car();

  bool res;
  if(foo)
    res = (v->number_of_wheels() == 2);
  else
    res = (v->number_of_wheels() == 4);
  assert(res);
  return 0;
}
\end{lstlisting}
\end{minipage}
\caption{C++ program using a simplified version of the UML diagram in
Fig.~\ref{figure:uml_diagram}. The program nondeterministically cast the derived
class to a base class. The goal is to check if the correct ${\tt number\_of\_wheels()}$
is called, from the base class.}
\label{figure:inheritance-example}
\end{figure}

Consider the program in Fig.~\ref{figure:inheritance-example}, which contains
a simplified version of the class hierarchy presented in
Fig.~\ref{figure:uml_diagram}. In the program, a class ${\tt Vehicle}$ is base for two
classes, ${\tt Motorcycle}$ and ${\tt Car}$. The class ${\tt Vehicle}$ defines a pure virtual method
${\tt number\_of\_wheel()}$, and both classes ${\tt Motorcycle}$ and ${\tt Car}$ implement the
method, returning 2 and 4, respectively. The program creates an instance of
${\tt Motorcycle}$ or ${\tt Car}$, depending on a nondeterministic choice, and assigns the
instance to a ${\tt Vehicle}$ pointer object ${\tt v}$. Finally, through the
polymorphic object ${\tt v}$, the program calls ${\tt number\_of\_wheel()}$
and checks the returned value. We omit a call to ${\tt delete}$ (that would free the pointer ${\tt v}$)
to simplify the GOTO instructions.

Fig.~\ref{figure:inheritance-goto} shows the GOTO program (resulted from the type-checking phase) generated for the
program in Fig.~\ref{figure:inheritance-example}. Note that, when building the
polymorphic object ${\tt v}$, the ${\tt vtable}$'s pointer for the
method ${\tt number\_of\_wheel()}$ is first assigned with
a pointer to the method ${\tt number\_of\_wheel()}$ in class ${\tt Vehicle}$
(see lines~\ref{veh_constr1} and~\ref{veh_constr2} in Fig.~\ref{figure:inheritance-goto}); this happens because the
constructor for both ${\tt Car}$ and ${\tt Motorcycle}$ first call the base constructor in the
original program (see lines~\ref{motor_constr} and~\ref{car_constr} in Fig.~\ref{figure:inheritance-example}). They are then
assigned the correct method address (see lines~\ref{motor_constr_goto}
and~\ref{car_constr_goto} in Fig.~\ref{figure:inheritance-goto}) in the constructors of the derived classes, \textup{i.e.}, ${\tt Motorcycle}$ and ${\tt Car}$, respectively.

In the SSA form shown in Fig.~\ref{figure:inheritance-SSA},
every branch creates a separate variable, which are
then combined when the control-flow merges. In Fig.~\ref{figure:inheritance-SSA}, we generate two
branches (i.e., ${\tt v1}$ and ${\tt v2}$) and a $\phi$-node (i.e., ${\tt v3}$)
to merge both branches. For instance, the variable ${\tt v1}$
represents the branch, where the polymorphic variable ${\tt v}$ gets assigned
an object of type ${\tt Motorcycle}$, while ${\tt v2}$ represents the branch, where
${\tt v}$ gets assigned an object of type ${\tt Car}$. They are then merged into
${\tt v3}$, depending on the initial nondeterministic choice (see
line~\ref{code-example:no-sideeffect} in Fig.~\ref{figure:inheritance-SSA}). There exists no
side-effect in the SSA form, %as the result of the ${\tt number\_of\_wheels()}$ is propagated.
as it can use the correct definition of ${\tt number\_of\_wheels()}$ in the $\phi$-node. The type-checker does all the heavy lifting.
   \begin{figure}[h!]
   \centering
    \subfloat[{GOTO instructions.}\label{figure:inheritance-goto}]{%
      \input{goto-instructions-poly.tex}
    }
    \hspace*{0.5cm}%
    \subfloat[{SSA form.}\label{figure:inheritance-SSA}]{%
      \input{SSA-form-poly.tex}
    }
    \caption{Internal representations of the program in Fig.~\ref{figure:inheritance-example}.}
    \label{figure:inheritance}
  \end{figure}

%--------------------------------------------
\section{C++ Operational Model}
\label{sec:cpp-operational-model}
%--------------------------------------------

The C++ programming language offers a collection of libraries, called STL, to provide most of the functionalities required by a programmer~\cite{ISO14882:2003}. However, the direct inclusion of the STL into the verification process over-complicates the analysis of C++ programs, as it contains code fragments not relevant for verification (\textup{e.g.}, optimized assembly code)~\cite{STVR2017, Ramalho13}. Its implementation is based on a pointer structure that degrades the verification performance~\cite{Blanc07}.
In particular, existing BMC tools adopt two different memory models: a \textit{fully byte-precise}~\cite{MerzFS12} or an \textit{object-based}~\cite{CBMC:2012,tacas14} memory model. Note that BMC tools reduce bounded program traces to a decidable fragment of first-order logic, which requires us to eliminate pointers in the model checker. They use static analysis to approximate each pointer variable the set of data objects (i.e., memory chunks) at which it might point at some stage in the program execution. For a \textit{fully byte-precise} memory model, BMC tools treat all memory as a single byte array, upon which all pointer accesses are decomposed into byte operations. This can lead to performance problems due to the repeated updates to the memory array that needs to be reflected in the SMT formula. For an \textit{object-based} memory model, this approach's performance suffers if pointer offsets cannot be statically determined, e.g., if a program reads a byte from an arbitrary offset into a structure. The resulting SMT formula is large and unwieldy, and its construction is error-prone.

To reduce verification complexity, ESBMC uses an abstract representation of the STL, called the C++ Operational Model (COM), which adds function contracts~\cite{cppcontracts2015} (i.e., pre- and post-conditions) to all STL function/method calls. Thus, all those function contracts are verified by ESBMC. The purpose of the verification is to check whether a given program uses correctly STL without hitting a bogus state (e.g., calling {\tt vector::operator[]} with an out-of-range parameter leads to undefined behavior).  A similar technique, proposed by Blanc et al.~\cite{Blanc07}, has been used to verify preconditions on programs. However, ESBMC extends that approach by also checking the post-conditions, which improves its effectiveness, as shown in our experimental evaluation (\textup{cf.}, Section~\ref{sec:experimental-evaluation}).

Fig.~\ref{figure:snippet} shows a code snippet considered as the best-accepted answer for a Stack Overflow question\footnote{Available at \url{https://stackoverflow.com/questions/41028862}.}. Nevertheless, line 10 could lead to an out-of-bound violation (CWE-125 vulnerability)~\cite{verdi2019empirical}. ESBMC detects the erroneous state through the operational model for {\tt vector::operator[]} (see Fig.~\ref{figure:operator}), which contains an assertion to check for out-of-bound accesses. The model also keeps track of the values stored in the container using a buffer ({\tt buf}), so it also guarantees the post-condition for the operator, i.e., return a reference to the element at specified location {\tt i}.
   \begin{figure}[h!]
   \centering
    \subfloat[{Code snippet.}\label{figure:snippet}]{%
      \input{stackoverflow-example.tex}
    }
    \hspace*{0.5cm}%
    \subfloat[{Operational model for {\tt vector::operator[]}.}\label{figure:operator}]{%
      \input{COM-example.tex}
    }
    \caption{Example from Stack Overflow (best accepted answer) that contains improper input validation (CWE-20) and out-of-bounds read (CWE-125) vulnerabilities.}
    \label{figure:om-example}
  \end{figure}
\begin{table*}[ht!]
\renewcommand\arraystretch{1.2}
\setlength{\tabcolsep}{1pt}
\begin{center} {\scriptsize%\small
\caption{Overview of the C++ Operational Model.}\label{table:overview-com}
\begin{tabular}{| >{\centering}m{1.5cm}| >{\centering}m{1.9cm} | >{\centering}m{1.7cm} | >{\centering}m{1.5cm} | >{\centering}m{1.8cm} | >{\centering}m{1.5cm} | >{\centering}m{1.2cm} | c |}
\hline
   \multicolumn{8}{|c|}{Standard C++03 Libraries -- Operational Model} \\
   \cline{1-8}
   C Standard Libraries & General & Streams Input/Output & Containers & Language Support & Numeric & Strings & Localization \\
   \hline
   \hline
   ${\tt cassert}$  	&  ${\tt memory}$      & ${\tt ios}$  		& ${\tt bitset}$  		& ${\tt exception}$    & ${\tt complex}$    	& ${\tt string}$  & ${\tt locale}$  \\ % OK
   ${\tt cctype}$  	&  ${\tt stdexcept}$  	& ${\tt iomanip}$  	& ${\tt deque}$  	& ${\tt limits}$    	& ${\tt random}$    	&   			&			\\ % OK
   ${\tt cerrno}$  	&  ${\tt utility}$  	        & ${\tt iosfwd}$  	& ${\tt list}$  		& ${\tt new}$    		& ${\tt valarray}$    	&   			&			\\ % OK
   ${\tt cfloat}$  		&  ${\tt functional}$   & ${\tt iostream}$  	& ${\tt map}$  		& ${\tt typeinfo}$    	& ${\tt numeric}$    	&   			&			\\ % OK
   ${\tt ciso646}$  	&    	                         & ${\tt istream}$  	& ${\tt multimap}$  	&     				&     				&   			&			\\ % OK
   ${\tt climits}$  	&    		                 & ${\tt ostream}$  	& ${\tt set}$  		&     				&     				&   			&			\\ % OK
   ${\tt clocale}$  	&   				& ${\tt streambuf}$  	& ${\tt multiset}$	&     				&     				&   			&			\\ % OK
   ${\tt cmath}$   	&    				& ${\tt sstream}$  	& ${\tt vector}$  	&     				&     				&   			&			\\ % OK
   ${\tt complex}$  	&    				& ${\tt fstream}$  	& ${\tt stack}$ 		&     				&     				&   			&			\\ % OK
   ${\tt csetjmp}$  	&    				&   				& ${\tt queue}$ 	        &     				&     				&   			&			\\ % OK
   ${\tt csignal}$  	&    				&   				& ${\tt algorithm}$	&     				&     				&   			&			\\ % OK
   ${\tt cstdarg}$  	&    				&   				& ${\tt iterator}$	        &     				&     				&   			&			\\ % OK
   ${\tt cstddef}$  	&    				&   				&                        	&     				&     				&   			&			\\ % OK
   ${\tt cstdio}$  		&    				&   				&   				&     				&     				&   			&			\\ % OK
   ${\tt cstdlib}$  	&    				&   				&   				&     				&     				&   			&			\\ % OK
   ${\tt cstring}$  	&    				&   				&   				&     				&     				&   			&			\\ % OK
   ${\tt ctime}$  		&    				&   				&   				&     				&     				&   			&			\\ % OK
\hline
\hline
\end{tabular} }
\end{center}
\end{table*}

Our COM mimics the structure of the STL, as shown in Table~\ref{table:overview-com}. All ANSI-C libraries are natively supported by ESBMC, as described by Cordeiro et al.~\cite{CordeiroFM12}. For all libraries under categories \texttt{General}, \texttt{Language Support}, \texttt{Numeric}, and \texttt{Localization}, COM adds pre-conditions extracted directly from documentation~\cite{ISO14882:2003}, specifically designed to detect memory-safety violations (e.g., nullness and out-of-bounds checks).

One of the challenges of modeling COM is the support for containers, strings, and streams, which requires the injection of pre- and post-conditions to check for functional properties correctly, as shown in the example illustrated in Fig.~\ref{figure:operator} (cf. the pre-conditions in lines 4-5). In this specific example, we check the \texttt{vector} upper and lower bounds before retrieving its content to detect an out-of-bounds read in line 10 of Fig.~\ref{figure:snippet}. COM models sequential and associative containers along with their iterators.  In particular, libraries ${\tt list}$, ${\tt bitset}$, ${\tt deque}$, ${\tt vector}$, ${\tt stack}$, and ${\tt queue}$ belong to the sequential group, while libraries ${\tt map}$, ${\tt multimap}$, ${\tt set}$, and ${\tt multiset}$ belong to the associative group. COM models strings and streams objects as arrays of bytes to properly encode them using the theory of arrays ({\it cf.}, Section~\ref{smt-theories}); therefore, ${\tt string}$ and all Stream I/O libraries also belong to the sequential group.

%--------------------------------------------
\subsection{Core Language}
\label{subsubsec:language}
%--------------------------------------------

The gist of COM enables ESBMC to encode features of standard containers, strings, and streams using the theory of arrays $\cal T_{A}$. To properly formalize the verification of our model, we extend the previous core container language presented by Ramalho \textup{et al.}~\cite{Ramalho13} to include a representation for keys, which allows us to reason about associative containers as well. The core language defines the syntactic domains values $\mathit{V}$, keys $\mathit{K}$, iterators $\mathit{I}$, pointers $\mathit{P}$, container $\mathit{C}$ and integers $\mathbb{N}$ as follows,
\begin{equation*}
\label{ccl-fig}
\small
\begin{array}{r@{\:\:}r@{\:\:}l}
  V  & := & v \:|\: \mathit{*i_v}
\\[0.5ex]
  K  & := & k \:|\: \mathit{*i_k}
\\[0.5ex]
   \mathit{I} & := &
     i \:|\: \mathit{C.insert(\mathit{I}, \mathit{V})} \:|\: \mathit{C.insert(\mathit{K}, \mathit{V})}
\\  &  &
     ~~~\:\mathit{C.search(\mathit{K})} \:|\: \mathit{C.search(\mathit{V})}
\\  &  &
     ~~~\:\mathit{C.erase(\mathit{I})}
\\[0.5ex]
   P  & := &
     p \:|\: P (+ \:|\: - ) P \:|\: c_v  \:|\: c_k \:|\: i_v  \:|\: i_k
\\[0.5ex]
   C  & := &
     c
\\[0.5ex]
  \mathbb{N}  & := & n \:|\: \mathbb{N} (+ \:|\: * | \ldots) \mathbb{N} \:|\: size \:|\: pos
  \end{array}
\end{equation*}

Here $v$, $k$, $p$, $i$, $c$ and $n$ are classes of variables of type $\mathit{V}$, $\mathit{K}$, $\mathit{P}$, $\mathit{I}$, $\mathit{C}$ and $\mathbb{N}$, respectively. For iterators, we use the notation $\mathit{*i_v}$ to denote the value stored in the memory location $i_v$. Based on such domains, we also define $P (+ \:|\: - ) P$ as valid pointer operations and $\mathbb{N} (+ \:|\: * | \ldots) \mathbb{N}$ as valid integer operations. Each operation shown in the core container syntax (\textup{e.g.}, $C.insert(I,V)$) is explained in Sections \ref{subsubsec:sequential} and \ref{subsubsec:associative}.

All methods from the sequential and associative groups can be expressed as combinations/variations of three main operations: insertion ($\mathit{C.insert(\mathit{I}, \mathit{V})}$), deletion ($\mathit{C.erase(\mathit{I})}$), and search ($\mathit{C.search(\mathit{V})}$). Each operation is described in our model as a Hoare triple $\{\cal P\} \: C \: \{\cal Q\}$ that represents the function contract scheme implemented by COM. Normally all side-effects would be stated in the post-condition $\cal Q$ for verification. However,  as part of the SSA transformation, side effects on iterators and containers are made explicit. Operations return new iterators and containers with the same contents, except for the fields that have just been updated. Thus, the translation function $\cal C$ contains primed variables ({\it e.g.}, $c'$ and $i'$) to represent the state of model variables after the respective operation. Finally, all models take advantage of {\it memcpy pattern} through lambda terms~\cite{boolector2015}, which enables us to describe array operations over multiple indices on a clear and concise manner ({cf.}, Section~\ref{smt-theories}).

%--------------------------------------------
\subsection{Sequential Containers}
\label{subsubsec:sequential}
%--------------------------------------------

\begin{wrapfigure}[12]{r}{7cm}
\centering
%\vspace{-12pt}
\includegraphics[scale=0.4]{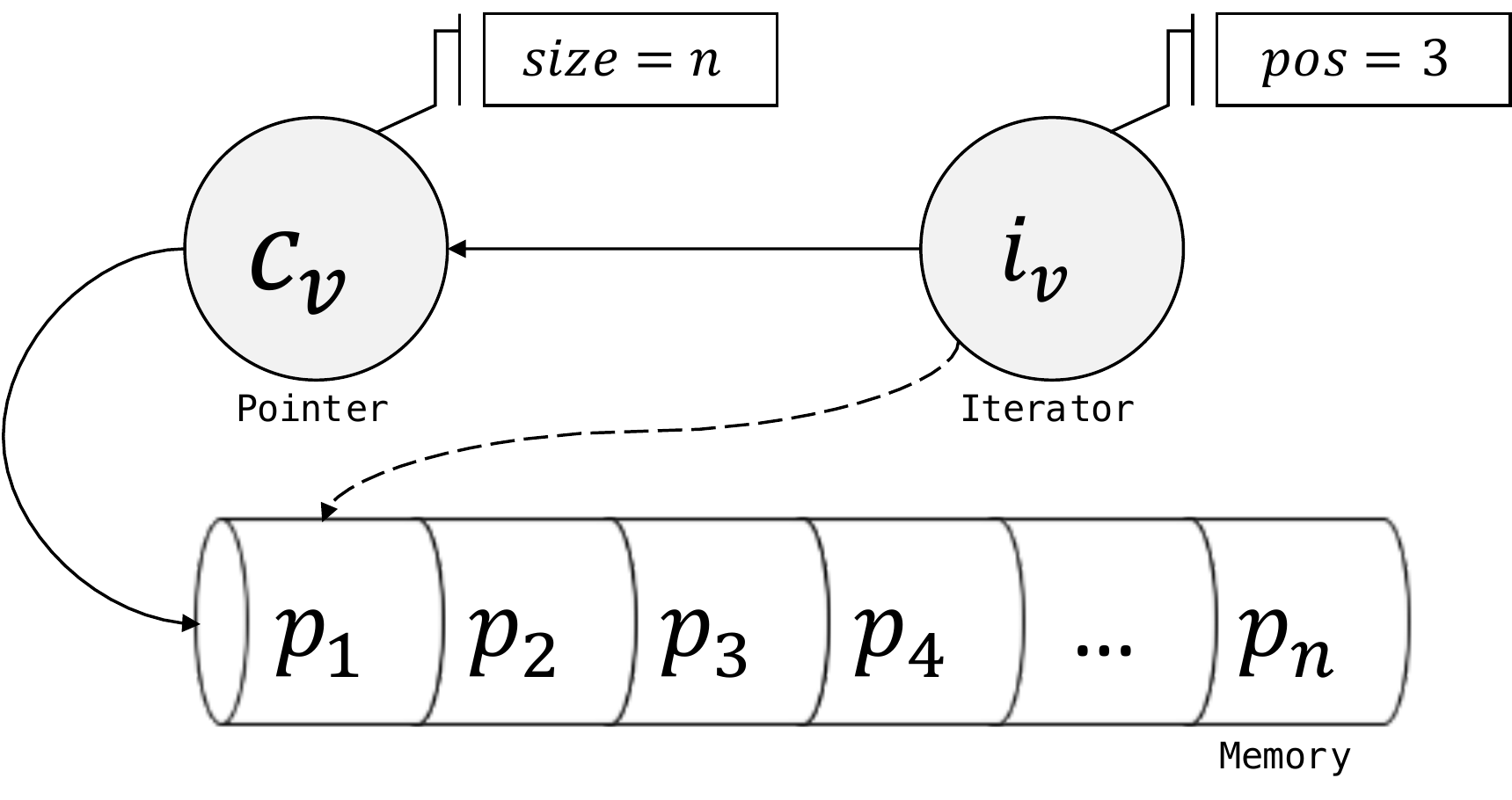}
\caption{Abstraction for sequential containers.}
\label{figure:stl-iterator}
\end{wrapfigure}

Sequential containers are built into a structure to
store elements with a sequential order~\cite{Deitel}. In our model, a sequential container
$\mathit{c}$ consists of a pointer $c_v$ that points to a valid memory location
and an integer $size$ that stores
the number of elements in the container. Similarly, an iterator $i$ is modeled
using two variables: an integer $i_{pos}$,
which contains the index value of the container pointed by the iterator and a
pointer $i_v$, which points to the memory location referred by the iterator.
In our model, the defined notation $\mathit{*i}$ is equivalent to $select(i_v, i_{pos})$.
Fig.~\ref{figure:stl-iterator} gives an overview of
our abstraction for all sequential containers.

The statement $\mathit{c.insert}(i,v)$ becomes $(c',i')=\mathit{c.insert}(i,v)$ increases the container size, move all elements from position $i.pos$ one memory unit forward, and then insert $v$ into the specified position. Therefore\footnote{Note that SMT theories only have a single equality predicate (for each sort). However, here we use the notation ``:='' to indicate an assignment of nested equality predicates on the right-hand side of the formula.},
\begin{equation}
\small
\label{sequential-c1}
\begin{array}{lll}
\multicolumn{3}{l}{{\cal C}((c',i')=c\mathit{.insert}(i, v)):=} \\
  & \multicolumn{2}{l}{c'.size = c.size + 1} \\
  & \multicolumn{2}{l}{\wedge \: {memcpy}(c.c_v, \: c'.c_v, \: i.pos, \: i.pos + 1, \: c.size - i.pos)} \\
  & \multicolumn{2}{l}{\wedge \: store(c'.c_v,\: i.pos,\: v)} \\
\end{array}
\end{equation}

\noindent that induces the following pre- and post-conditions,
\begin{equation}
\small
\begin{array}{lll}
\multicolumn{3}{l}{{\cal P}((c',i')=c\mathit{.insert}(i, v)):=} \\
  & \multicolumn{2}{l}{v \: \neq \:  null} \\
  & \multicolumn{2}{l}{\wedge\: c.c_v \: \neq \:  null} \\
  & \multicolumn{2}{l}{\wedge\: i.i_v \: \neq \:  null} \\
  & \multicolumn{2}{l}{\wedge\: 0 \: \leq \: i.pos \: < \: c'.size} \\
\end{array}
\end{equation}

\begin{equation}
\small
\begin{array}{lll}
\multicolumn{3}{l}{{\cal Q}((c',i')=c\mathit{.insert}(i, v)):=} \\
  & \multicolumn{2}{l}{select(i'.i_v,\: i'.pos) \: = \: v} \\
  & \multicolumn{2}{l}{\wedge\:i'.i_v = c'.c_v} \\
  & \multicolumn{2}{l}{\wedge\:i'.pos = i.pos}
\end{array}
\end{equation}

\noindent where $null$ represents an uninitialized pointer/object.
Thus, we define as pre-conditions ${\cal P}$ that $v$ and $i$ can not be uninitialized objects
as well as $i.pos$ must be within $c'.c_v$ bounds; similarly, we define as
post-conditions ${\cal Q}$ that $v$ was correctly inserted in the position specified by $i$ as
well as $c'.c_v$ and $i'.i_v$ are equivalent, \textup{i.e.}, both point to the same
memory location. Importantly, we implement the memory model for containers
essentially as arrays, therefore, the range to select elements from memory varies
from $0$ to $c.size-1$. Furthermore, the main effect of the \textit{insert} method is mainly captured
by Eq.~\eqref{sequential-c1} that describes the contents of the container array $c'.c_v$
after the insertion in terms of update operations to the container array
$c.c_v$ before the insertion.

The erase method works similarly to the insert method. It uses iterator positions, integer values, and pointers, but it does not use values since the exclusion is made by a given position, regardless of the value. It also returns an iterator position (\textup{i.e.}, $i'$), pointing to the position immediately after the erased part of the container~\cite{ISO14882:2003}. Therefore,
\begin{equation}
\small
\label{sequential-c2}
\begin{array}{lll}
\multicolumn{3}{l}{{\cal C}((c',i')=c\mathit{.erase}(i)):=} \\
  & \multicolumn{2}{l}{{memcpy}(c.c_v, \: c'.c_v, \: i.pos + 1, \: i.pos, \: c.size - (i.pos + 1))} \\
  & \multicolumn{2}{l}{\wedge\: c'.size = c.size - 1} \\
  & \multicolumn{2}{l}{\wedge\: i'.pos = i.pos + 1}
\end{array}
\end{equation}

\noindent that induces the following pre- and post-conditions,
\begin{equation}
\small
\label{sequential-p2}
\begin{array}{lll}
\multicolumn{3}{l}{{\cal P}((c',i')=c\mathit{.erase}(i)):=} \\
  & \multicolumn{2}{l}{i.i_v \: \neq \:  null} \\
  & \multicolumn{2}{l}{\wedge\: c.c_v \neq null} \\
  & \multicolumn{2}{l}{\wedge\: 0 \: \leq \: i.pos \: < \: c.size} \\
  & \multicolumn{2}{l}{\wedge\: c.size \neq 0 \Rightarrow c.c_v \neq null}
\end{array}
\end{equation}
\begin{equation}
\small
\label{sequential-q2}
\begin{array}{lll}
\multicolumn{3}{l}{{\cal Q}((c',i')=c\mathit{.erase}(i)):=} \\
  & \multicolumn{2}{l}{select(c'.c_v,\: i'.pos) \: = \: select(c.c_v,\: i.pos+1)} \\
  & \multicolumn{2}{l}{\wedge\: i'.i_v = {c'}.c_v} \\
\end{array}
\end{equation}

\noindent where we assume as pre-conditions ${\cal P}$ that $i$ must be a
valid iterator pointing to a position within the bounds of array $c.c_v$ and $c$
must be non-empty; similarly, we assume as post-conditions ${\cal Q}$ that $i'$ must point
to the element immediately after the erased one and $c'.c_v$ and $i'.i_v$ point
to the same memory location. Finally, a container $c$ with a call \textit{c.search}($v$)
performs a search for an element $v$ in the container. Then, if such an element
is found, it returns an iterator that points to the respective element;
otherwise, it returns an iterator that points to the position immediately after
the last container's element (i.e., $\mathit{select(c'.c_v, \: c'.size)}$). Hence,
\begin{equation}
\small
\label{sequential-c3}
\begin{array}{lll}
\multicolumn{3}{l}{{\cal C}((c',i')=c\mathit{.search}(v)):=}\\
  & \multicolumn{2}{l}{ite\big(c.size = 0,}\\
  & \multicolumn{2}{l}{~~~~~~i'.pos = c.size,}\\
  & \multicolumn{2}{l}{~~~~~~ite(select(c.c_v, 0) = v,}\\
  & \multicolumn{2}{l}{~~~~~~~~~~~~i'.pos = 0,}\\
  & \multicolumn{2}{l}{~~~~~~~~~~~~...}\\
  & \multicolumn{2}{l}{~~~~~~~~~~~~ite(select(c.c_v, c.size - 1) = v,}\\
  & \multicolumn{2}{l}{~~~~~~~~~~~~~~~~~~i'.pos = c.size -1,}\\
  & \multicolumn{2}{l}{~~~~~~~~~~~~~~~~~~i'.pos = c.size) \: ... \: )\big)}
\end{array}
\end{equation}

\noindent that induces the following pre- and post-conditions,
\begin{equation}
\small
\label{sequential-p3}
\begin{array}{lll}
\multicolumn{3}{l}{{\cal P}((c',i')=c\mathit{.search}(v)):=} \\
  & \multicolumn{2}{l}{v \: \neq \:  null}
\end{array}
\end{equation}
\begin{equation}
\small
\label{sequential-q3}
\begin{array}{lll}
\multicolumn{3}{l}{{\cal Q}((c',i')=c\mathit{.search}(v)):=} \\
  & \multicolumn{2}{l}{c'.c_v = c.c_v}\\
  & \multicolumn{2}{l}{\wedge\: c'.size = c.size}\\
  & \multicolumn{2}{l}{\wedge\: i'.i_v \neq c'.c_v} \\
  & \multicolumn{2}{l}{\wedge\: ite\big(select(i'.i_v,\: i'.pos) \: = \: select(c'.c_v, i'.pos),} \\
  & \multicolumn{2}{l}{~~~~~~~~~select(i'.i_v, \: i'.pos) = v,} \\
  & \multicolumn{2}{l}{~~~~~~~~~select(i'.i_v, \: i'.pos) = select(c'.c_v, \: c'.size)\big)}
\end{array}
\end{equation}

\noindent where we assume as pre-conditions ${\cal P}$ that
$v$ and $c$ can not be an uninitialized objects;
similarly, we assume as post-conditions ${\cal Q}$ that $c'$
is equivalent to its previous state $c$, $c'.c_v$ and $i'.i_v$ point
to the same memory location, and $i'$ must point to the found element or to
$\mathit{select(c'.c_v, \: c'.size)}$.

%--------------------------------------------
\subsection{Associative Containers}
\label{subsubsec:associative}
%--------------------------------------------
%
\begin{wrapfigure}[20]{r}{7.5cm}
\centering
%\vspace{-12pt}
\includegraphics[scale=0.4]{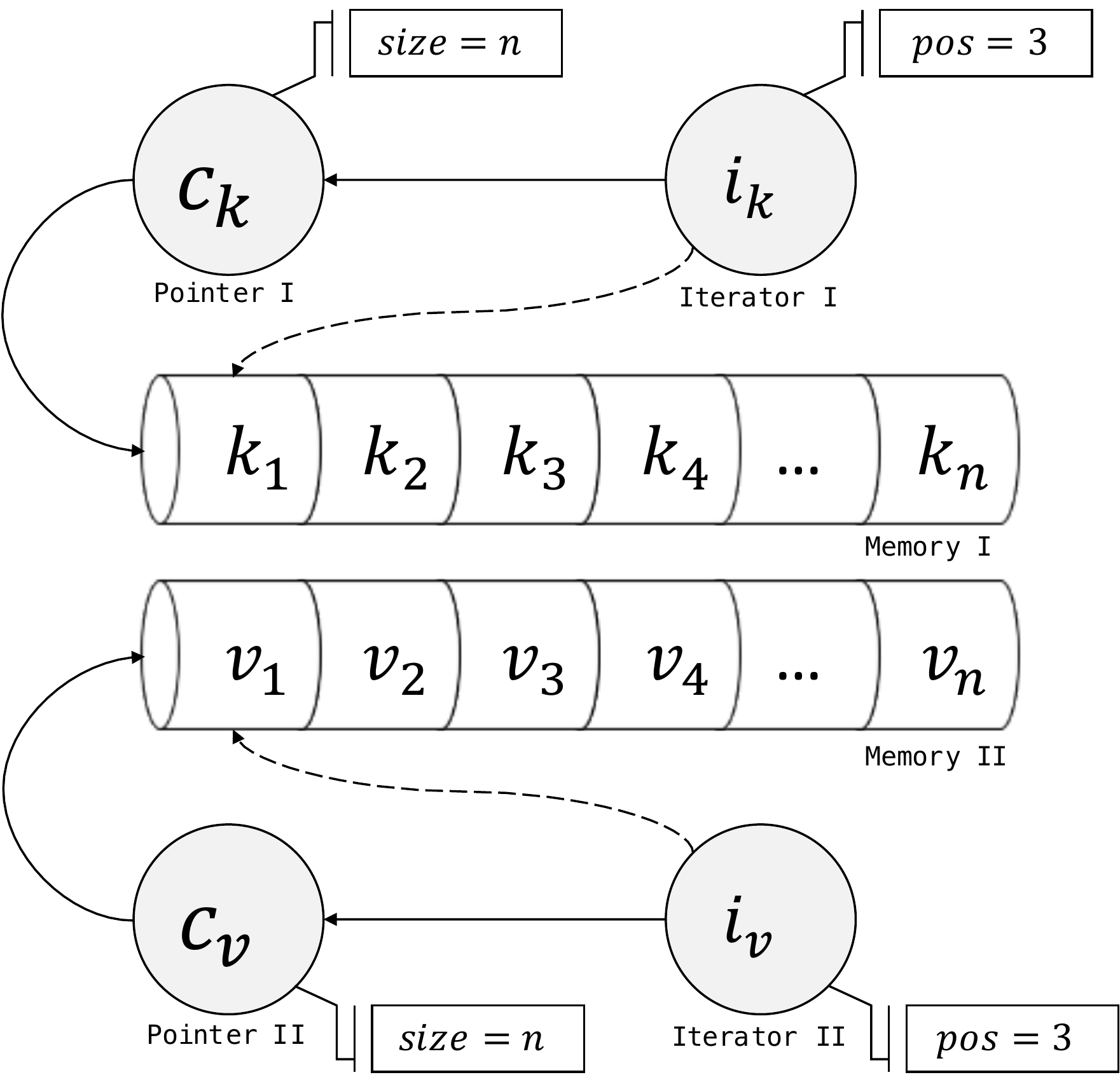}
\caption{Abstraction for associative containers.}
\label{figure:stl-asso-iterator}
\end{wrapfigure}

Associative containers consist of elements with a key $k$ and a value $v$, where
each value is associated with a unique key. All elements are internally sorted by
their keys based on a strict weak ordering rule~\cite{ISO14882:2003}.
In our model, an associative container $\mathit{c}$ consists of a pointer
$c_v$, for the container's values, a pointer $c_k$,
for the container's keys, and an integer $size$, for the container's size.
Fig.~\ref{figure:stl-asso-iterator} gives an overview of
our abstraction for all associative containers.
The relationship between $c_k$ and $c_v$ is established by an index, thus, an element in a
given position $n$ in $c_k$ (\textup{i.e.}, $select(c.c_k, \: n)$) is the key
associated with the value in the same position $n$ in $c_v$ (\textup{i.e.},
$select(c.c_v, \: n)$). Similarly, iterators for associative containers consist
of a pointer $\mathit{i_k}$ that points to the same memory location
as $c_k$, a pointer $\mathit{i_v}$ that points to the same memory location as
$c_v$, and an integer $\mathit{i_{pos}}$ that indexes both $\mathit{i_k}$ and $\mathit{i_v}$.
All operations for associative containers can be expressed
as a simplified variation of the three main ones, \textup{i.e.}, insertion
($\mathit{C.insert(\mathit{K}, \mathit{V})}$), deletion
($\mathit{C.erase(\mathit{I})}$), and search ($\mathit{C.search(\mathit{K})}$).

The order of keys matters in the insertion operation for associative containers. Therefore, given a container $c$, the method calls $c.insert(k,v)$ inserts the value $v$ associated with the key $k$ into the right order (i.e., obeying a strict weak ordering rule). Here, we use the operator $\prec$ to represent precedence; thus, $x \prec y$ means $x$ precedes $y$. The insertion returns an iterator that points to the inserted position. However, if $k$ exists, the insertion is not performed and the method returns an iterator that points to the existing element. We checked for three cases, which correspond to each $ite$ condition: {\it (i)} the empty case first, then {\it (ii)} we check whether each position contains a corresponding key or {\it (iii)} if we should insert the value based on its precedence. Thus,
\begin{equation}
\small
\label{associative-c1}
\begin{array}{lll}
\multicolumn{3}{l}{{\cal C}((c',i')=c\mathit{.insert}(k,v)):=}\\
  & \multicolumn{2}{l}{ite\big(c.size = 0,}\\
  & \multicolumn{2}{l}{~~~~~~i'.pos = 0}\\
  & \multicolumn{2}{l}{~~~~~\wedge \: {store}(c'.c_k,\: i'.pos,\: k)}\\
  & \multicolumn{2}{l}{~~~~~\wedge \: {store}(c'.c_v,\: i'.pos,\: v)}\\
  & \multicolumn{2}{l}{~~~~~\wedge \: c'.size = c.size + 1,}\\
  & \multicolumn{2}{l}{~~~~~~ite(select(c.c_k, 0) = k,}\\
  & \multicolumn{2}{l}{~~~~~~~~~~~~i'.pos = 0,}\\
  & \multicolumn{2}{l}{~~~~~~~~~~~~ite(k \prec select(c.c_k, 0),} \\
  & \multicolumn{2}{l}{~~~~~~~~~~~~~~~~~~i'.pos = 0}\\
  & \multicolumn{2}{l}{~~~~~~~~~~~~~~~~~\wedge \: {memcpy}(c.c_k, \: c'.c_k, \: i'.pos, \: i'.pos + 1, \: c.size - i'.pos)} \\
  & \multicolumn{2}{l}{~~~~~~~~~~~~~~~~~\wedge \: {store}(c'.c_k,\: i'.pos,\: k)} \\
  & \multicolumn{2}{l}{~~~~~~~~~~~~~~~~~\wedge \: {memcpy}(c.c_v, \: c'.c_v, \: i'.pos, \: i'.pos + 1, \: c.size - i'.pos)} \\
  & \multicolumn{2}{l}{~~~~~~~~~~~~~~~~~\wedge \: {store}(c'.c_v,\: i'.pos,\: v)} \\
  & \multicolumn{2}{l}{~~~~~~~~~~~~~~~~~\wedge \: c'.size = c.size + 1,}\\
  & \multicolumn{2}{l}{~~~~~~~~~~~~~~~~~~\cdots}\\
  & \multicolumn{2}{l}{~~~~~~~~~~~~~~~~~~ite(select(c.c_k, c.size - 1) = k,}\\
  & \multicolumn{2}{l}{~~~~~~~~~~~~~~~~~~~~~~~~i'.pos = c.size - 1,}\\
  & \multicolumn{2}{l}{~~~~~~~~~~~~~~~~~~~~~~~~ite(k \prec select(c.c_k, \: c.size - 1), i'.pos = c.size - 1, i'.pos = c.size)} \\
  & \multicolumn{2}{l}{~~~~~~~~~~~~~~~~~~~~~~~\wedge \: {memcpy}(c.c_k, \: c'.c_k, \: i'.pos, \: i'.pos + 1, \: c.size - i'.pos)} \\
  & \multicolumn{2}{l}{~~~~~~~~~~~~~~~~~~~~~~~\wedge \: {store}(c'.c_k,\: i'.pos,\: k)} \\
  & \multicolumn{2}{l}{~~~~~~~~~~~~~~~~~~~~~~~\wedge \: {memcpy}(c.c_v, \: c'.c_v, \: i'.pos, \: i'.pos + 1, \: c.size - i'.pos)} \\
  & \multicolumn{2}{l}{~~~~~~~~~~~~~~~~~~~~~~~\wedge \: {store}(c'.c_v,\: i'.pos,\: v)} \\
  & \multicolumn{2}{l}{~~~~~~~~~~~~~~~~~~~~~~~\wedge \: c'.size = c.size + 1) \: \cdots \:))\big)}\\
\end{array}
\end{equation}

\noindent that induces the following pre- and post-conditions,
\begin{equation}
\small
\begin{array}{lll}
\multicolumn{3}{l}{{\cal P}((c',i')=c\mathit{.insert}(k, v)):=} \\
  & \multicolumn{2}{l}{k \: \neq \:  null} \\
  & \multicolumn{2}{l}{\wedge\: v \: \neq \:  null} \\
  & \multicolumn{2}{l}{\wedge\: \big(\bigwedge \limits_{j=0}^{c.size -2} select(c.c_k, j) \prec selct(c.c_k, j + 1)\big)}
\end{array}
\end{equation}
\begin{equation}
\small
\begin{array}{lll}
\multicolumn{3}{l}{{\cal Q}((c',i')=c\mathit{.insert}(k, v)):=} \\
  & \multicolumn{2}{l}{i'.i_k = c'.c_k} \\
  & \multicolumn{2}{l}{\wedge\: i'.i_v = c'.c_v} \\
  & \multicolumn{2}{l}{\wedge\: \big(\bigwedge \limits_{j=0}^{c.size -1} select(c.c_k, j) \neq k\big) \Rightarrow c'.size = c.size + 1} \\
  & \multicolumn{2}{l}{\wedge\: \big(\bigwedge \limits_{j=1}^{c.size -1} select(c.c_k, j-1) \prec selct(c.c_k, j)\big)}
\end{array}
\end{equation}

\noindent where we assume as pre-conditions ${\cal P}$ that $v$ and $k$ must be initialized objects, as well as the order of elements, obey a strict weak ordering rule. Similarly, we assume as post-conditions ${\cal Q}$ that the iterator $i'$ will point to the container $c'$, and the strict weak ordering rule will be maintained. We also check whether the size of the container will grow if the key $k$ was not used before; however, this check is bypassed for containers that allow multiple keys.

Remove operations are represented by $c.erase$($i$), where $i$ is
an iterator that points to the element to be removed. Similarly to sequential
containers (\textit{cf.}, Section~\ref{subsubsec:sequential}), the model for
such operation basically shifts backwards all elements followed by that specific
position $i$. Thus,
\begin{equation}
\small
\label{associative-c2}
\begin{array}{lll}
\multicolumn{3}{l}{{\cal C}((c',i')=c\mathit{.erase}(i)):=}\\
  & \multicolumn{2}{l}{{memcpy}(c.c_k, \: c'.c_k, \: i.pos + 1, \: i.pos, \: c.size - (i.pos + 1))} \\
  & \multicolumn{2}{l}{\wedge \: {memcpy}(c.c_v, \: c'.c_v, \: i.pos + 1, \: i.pos, \: c.size - (i.pos + 1))} \\
  & \multicolumn{2}{l}{\wedge\: c'.size = c.size - 1}\\
  & \multicolumn{2}{l}{\wedge\: i'.pos = i.pos + 1}\\
\end{array}
\end{equation}

\noindent that induces the following pre- and post-conditions,
\begin{equation}
\small
\label{associative-p2}
\begin{array}{lll}
\multicolumn{3}{l}{{\cal P}((c',i')=c\mathit{.erase}(i)):=} \\
  & \multicolumn{2}{l}{i.i_k \: \neq \:  null} \\
  & \multicolumn{2}{l}{i.i_v \: \neq \:  null} \\
  & \multicolumn{2}{l}{\wedge\: 0 \: \leq \: i.pos \: < \: c.size} \\
  & \multicolumn{2}{l}{\wedge\: c.size \neq 0 \Rightarrow (c.c_k \neq null \wedge c.c_v \neq null)}
\end{array}
\end{equation}
\begin{equation}
\small
\label{associative-p3}
\begin{array}{lll}
\multicolumn{3}{l}{{\cal Q}((c',i')=c\mathit{.erase}(i)):=} \\
  & \multicolumn{2}{l}{i'.i_k = {c'}_k}\\
  & \multicolumn{2}{l}{\wedge\: i'.i_v = {c'}_v}\\
  & \multicolumn{2}{l}{\wedge\: select(c'.c_k,\: i'.pos) \: = \: select(c.c_k,\: i.pos+1)} \\
  & \multicolumn{2}{l}{\wedge\: select(c'.c_v,\: i'.pos) \: = \: select(c.c_v,\: i.pos+1)} \\
\end{array}
\end{equation}

\noindent which have similar properties as the ones held by the {\it erase} method from sequential containers, except that $i'.i_k$ must point to the position immediately after the erased one and the equivalency of $c'.c_k$ and $i'.i_k$. Finally, search operations over associative containers are modeled by a container $c$ with a method call \textit{c.search}($k$). Then, if an element with key $k$ is found, the method returns an iterator that points to the corresponding element; otherwise, it returns an iterator that points to the position immediately after the last container's element. Hence,
\begin{equation}
\small
\label{sequential-c4}
\begin{array}{lll}
\multicolumn{3}{l}{{\cal C}((c',i')=c\mathit{.search}(k)):=}\\
  & \multicolumn{2}{l}{ite\big(c.size = 0,}\\
  & \multicolumn{2}{l}{~~~~~~i'.pos = c.size,}\\
  & \multicolumn{2}{l}{~~~~~~ite(select(c.c_k, 0) = k,}\\
  & \multicolumn{2}{l}{~~~~~~~~~~~~i'.pos = 0,}\\
  & \multicolumn{2}{l}{~~~~~~~~~~~~...}\\
  & \multicolumn{2}{l}{~~~~~~~~~~~~ite(select(c.c_k, c.size - 1) = k,}\\
  & \multicolumn{2}{l}{~~~~~~~~~~~~~~~~~~i'.pos = c.size -1,}\\
  & \multicolumn{2}{l}{~~~~~~~~~~~~~~~~~~i'.pos = c.size) \: ... \: )\big)}
\end{array}
\end{equation}

\noindent that induces the following pre- and post-conditions,
\begin{equation}
\small
\label{sequential-p4}
\begin{array}{lll}
\multicolumn{3}{l}{{\cal P}((c',i')=c\mathit{.search}(k)):=} \\
  & \multicolumn{2}{l}{k \: \neq \:  null}
\end{array}
\end{equation}
\begin{equation}
\small
\label{sequential-q4}
\begin{array}{lll}
\multicolumn{3}{l}{{\cal Q}((c',i')=c\mathit{.search}(v)):=} \\
  & \multicolumn{2}{l}{c'.c_k = c.c_k}\\
  & \multicolumn{2}{l}{\wedge\: c'.c_v = c.c_v}\\
  & \multicolumn{2}{l}{\wedge\: c'.size = c.size}\\
  & \multicolumn{2}{l}{\wedge\: i'.i_k \neq c'.c_k} \\
  & \multicolumn{2}{l}{\wedge\: i'.i_v \neq c'.c_v} \\
  & \multicolumn{2}{l}{\wedge\: ite\big(select(i'.i_k,\: i'.pos) \: = \: select(c'.c_k, i'.pos),} \\
  & \multicolumn{2}{l}{~~~~~~~~~select(i'.i_k, \: i'.pos) = k,} \\
  & \multicolumn{2}{l}{~~~~~~~~~select(i'.i_k, \: i'.pos) = select(c'.c_k, \: c'.size)\big)} \\
  & \multicolumn{2}{l}{\wedge\: ite\big(select(i'.i_v,\: i'.pos) \: = \: select(c'.c_v, i'.pos),} \\
  & \multicolumn{2}{l}{~~~~~~~~~select(i'.i_v, \: i'.pos) = v,} \\
  & \multicolumn{2}{l}{~~~~~~~~~select(i'.i_v, \: i'.pos) = select(c'.c_v, \: c'.size)\big)} \\
\end{array}
\end{equation}

\noindent that are also similar to the properties held by the
{\it search} operation from sequential containers, except that the search happens over keys.

%------------------------------------------------
\section{Exception Handling}
\label{sec:exception-handling}
%------------------------------------------------

Exceptions are unexpected circumstances that arise during the execution of a
program, \textup{e.g.}, runtime errors~\cite{Deitel}. In C++, the exception
handling is split into three (basic) elements: a \texttt{try} block, where
a thrown exception can be directed to a \texttt{catch} statement; a set of
\texttt{catch} statements, where a thrown exception can be handled; and a
\texttt{throw} statement that raises an exception.

To accurately define the verification of exception handling in C++, we formally define two syntactic domains, including exceptions $\mathit{E}$ and handlers $\mathit{H}$ as follows:
\begin{wrapfigure}[5]{r}{6cm}
\centering
\begin{equation*}
\label{exception-core-language}
\small
{\begin{array}{r@{\:\:}r@{\:\:}l}
  E  & := &
    \mathit{e} \:|\: \mathit{e_{[]}} \:|\: \mathit{e_{f()}} \:|\: \mathit{e_{*}} \:|\: \mathit{e_{null}}
\\[0.5ex]
  H  & := &
    \mathit{h} \:|\: \mathit{h_{[]}} \:|\: \mathit{h_{f()}} \:|\: \mathit{h_{*}} \:|\: \mathit{h_{v}} \:|\: \mathit{h_{...}}  \:|\: \mathit{h_{null}}
  \end{array}}
\end{equation*}
\end{wrapfigure}

In this context, $e$ and $h$ are classes of variables of type $\mathit{E}$ and $\mathit{H}$, respectively.
We use the notation $\mathit{e_{[]}}$ to denote a thrown exception of type array,
$\mathit{e_{f()}}$ is a thrown exception of type function, $\mathit{e_{*}}$ is a thrown exception of type pointer, and $\mathit{e_{null}}$ is an empty exception used to track when a \textit{throw} expression does not throw anything. Similarly, we use the notation $\mathit{h_{[]}}$ to denote a \texttt{catch} statement of type array,
$\mathit{h_{f()}}$ is a \texttt{catch} statement of type function, $\mathit{h_{*}}$ is a \texttt{catch} statement of type pointer, $\mathit{h_{v}}$ is a \texttt{catch} statement of type void pointer (i.e., ${\tt void *}$), $\mathit{h_{...}}$ is a \texttt{catch} statement of type ellipsis~\cite{ISO14882:2003}, and $\mathit{h_{null}}$ is an invalid \texttt{catch} statement used to track when a thrown exception does not have a valid handler.

Based on such domains, we must define a $2$-arity predicate $M(e, h)$, which evaluates whether the type of thrown exception $e$ is compatible with the type of a given handler $h$ as shown in Eq.~\eqref{match}. Furthermore, we declare the unary function $\zeta: H^* \longmapsto H$ that removes qualifiers \texttt{const}, \texttt{volatile}, and \texttt{restrict} from the type of a \texttt{catch} statement $c$. We also define the $2$-arity predicates unambiguous base $U(e, h)$ and implicit conversion $Q(e, h)$. On one hand, $U(e, h)$ determines whether the type of a \texttt{catch} statement $h$ is an unambiguous base~\cite{ISO14882:2003} for the type of a thrown exception $e$ as shown in Eq.~\eqref{unambiguous}. On the other hand, $Q(e, h)$ determines whether a thrown exception $e$ can be converted to the type of the \texttt{catch} statement $h$, either by qualification or standard pointer conversion~\cite{ISO14882:2003} as shown in Eq.~\eqref{conversion}.
\begin{equation}
\label{match}
\small
{  \begin{array}{ll}
  M(e, h) \eqdef \left\{
    \begin{array}{c l}
      \top, & $type \: of \:$ e $\: is \: matches \: to \: the \: type \: of\:$ h\\
      \bot, & $otherwise$
    \end{array}\right.
  \end{array}}
\end{equation}
\begin{equation}
\label{unambiguous}
\small
{  \begin{array}{ll}
  U(e, h) \eqdef \left\{
    \begin{array}{c l}
      \top, & c $\: is \: an \: unambiguous \: base \: of \:$ e\\
      \bot, & $otherwise$
    \end{array}\right.
  \end{array}}
\end{equation}
\begin{equation}
\label{conversion}
\small
{  \begin{array}{ll}
  Q(e, h) \eqdef \left\{
    \begin{array}{c l}
      \top, & e $\: can \: be \: implicit \: converted \: to \:$ h\\
      \bot, & $otherwise$
    \end{array}\right.
  \end{array}}
\end{equation}

The C++ language standard defines rules to connect
\texttt{throw} expressions and \texttt{catch} statements~\cite{ISO14882:2003},
which are all described in Table~\ref{table:rules-catch}.
Each rule represents a function $r_k: E \longmapsto H$ for $k = [1 \: .. \: 9]$,
where a thrown exception $e$ is mapped to a valid \texttt{catch} statement $h$.
ESBMC evaluates every thrown exception $e$ against all rules and all
\texttt{catch} statements in the program through the $(n+1)$-arity function
handler $\mathcal{H}$. As shown in Eq.~\eqref{rules}, after the evaluation of
all rules (i.e., $h_{r_1}, ..., h_{r_9}$), ESBMC returns the first handler
$h_{r_k}$ that matched the thrown exception $e$.
\begin{equation}
\label{rules}
\small
  {\begin{array}{l}
    \mathcal{H}(e, h_1, ..., h_n) := \\
    ~~~~~~~~~~~~~~h_{r_1} = r_1(e, h_1, ..., h_n)\\
    ~~~~~~~~~~~~~~\wedge \ldots  \\
    ~~~~~~~~~~~~~~\wedge \: h_{r_9} = r_9(e, h_1, ..., h_n) \\
    ~~~~~~~~~~~~~~\wedge \: ite(h_{r_1} \neq h_{null}, h_{r_1}, \\
    ~~~~~~~~~~~~~~~~~~~~ite(h_{r_2} \neq h_{null}, h_{r_2},\\
    ~~~~~~~~~~~~~~~~~~~~\ldots \\
    ~~~~~~~~~~~~~~~~~~~~ite(h_{r_9} \neq h_{null}, h_{r_9}, h_{null})\ldots)
  \end{array}}
\end{equation}

To support exception handling in ESBMC, we extended our GOTO conversion code and the symbolic engine. In the former, we had to define new instructions and model the throw expression as jumps. In the latter, we implemented the rules for throwing and catching exceptions, as shown in Table~\ref{table:rules-catch}, and the control flows for the unexpected and terminate handlers ({\it cf.}, Section~\ref{sec:terminate-and-unexpected}).
\begin{table*}[!ht]
\renewcommand\arraystretch{1.18}
\setlength{\tabcolsep}{4pt}
\caption{{Rules to connect \texttt{throw} expressions and \texttt{catch} blocks.}}
\begin{center} {\scriptsize
\begin{tabular}{| >{\centering\arraybackslash} m{1cm} | p{4cm} | >{\centering\arraybackslash} m{9cm} |}
\hline
{\textbf{Rule}}    & \multicolumn{1}{c|}{{\textbf{Behavior}}} & \multicolumn{1}{c|}{{\textbf{Formalization}}}  \\
\hline
{$r_1$} & {Catches an exception if the type of the thrown exception $e$ is equal to
the type of the \texttt{catch} $h$.} & {$ite(\exists h \cdot \: M(e, h), h_{r_1} = h, h_{r_1} = h_{null})$}\\
\hline
{$r_2$} & {Catches an exception if the type of the thrown exception $e$ is equal to
the type of the \texttt{catch} $h$, ignoring the qualifiers $const$,
$volatile$, and $restrict$.}  & {$ite(\exists h \cdot \: M(e, \zeta(h)), h_{r_2} = h, h_{r_2} = h_{null})$} \\
\hline
{$r_3$} & {Catches an exception if its type is a pointer of a given type $x$ and the type of the
thrown exception is an array of the same type $x$.} & {$ite(\exists h \cdot \: e = e_{[]} \wedge h = h_{*} \wedge M(e_{[]}, h_{*}), h_{r_3} = h_{*}, h_{r_3} = h_{null})$} \\
\hline
{$r_4$} & {Catches an exception if its type is a pointer to function that returns
a given type $x$ and the type of the thrown exception is a function that returns the same type
$x$.}  & {$ite(\exists h \cdot \: e = e_{f()} \wedge h = h_{f()} \wedge M(e_{f()}, h_{f()}), h_{r_4} = h_{f()}, h_{r_4} = h_{null})$} \\
\hline
{$r_5$} & {Catches an exception if its type is an unambiguous base type for the
type of the thrown exception.} & {$ite(\exists h \cdot \: U(e, h), h_{r_5} = h, h_{r_5} = h_{null})$}  \\
\hline
{$r_6$} & {Catches an exception if the type of the thrown exception $e$ can be
converted to the type of the \texttt{catch} $h$, either by qualification or standard
pointer conversion \cite{ISO14882:2003}.}  & {$ite(\exists h \cdot \: e = e_{*} \wedge h = h_{*} \wedge Q(e_{*}, h_{*}), h_{r_6} = h_{*}, h_{r_6} = h_{null})$} \\
\hline
{$r_7$} & {Catches an exception if its type is a void pointer $h_v$ and the type of the
thrown exception $e$ is a pointer of any given type.} & {$ite(\exists h \cdot \: e = e_{*} \wedge h = h_{v}, h_{r_7} = h_{v}, h_{r_7} = h_{null})$} \\
\hline
{$r_8$} & {Catches any thrown exception if its type is ellipsis.} & {$ite(\forall e \cdot \exists h \cdot \: h = h_{...}, h_{r_8} = h_{...}, h_{r_8} = h_{null})$} \\
\hline
{$r_9$} & {If the \texttt{throw} expression does not throw anything, it should re-throw
the last thrown exception $e_{-1}$, if it exists.} & $ \begin{array}{lll}
\multicolumn{3}{l}{{ite(e = e_{null} \wedge e_{-1} \neq e_{null},}} \\
  & \multicolumn{2}{l}{{~~~h_{r_1}' = r_1(e_{-1}, h_1, ..., h_n)}} \\
  & \multicolumn{2}{l}{{~~~\wedge\: \ldots}} \\
  & \multicolumn{2}{l}{{~~~\wedge\: h_{r_9}' = r_9(e_{-1}, h_1, ..., h_n),}} \\
  & \multicolumn{2}{l}{{~~~h_{r_9} = h_{null})}}
%  & \multicolumn{2}{l}{~~~~~~~~~~~~~~~~~~~{\wedge\: handling(h_{r_1}', \ldots, h_{r_9}')}} \\
\end{array}$ \\
\hline
\end{tabular} }
\end{center}
\label{table:rules-catch}
\end{table*}

The GOTO conversion slightly modifies the exception handling blocks $H$. The following instructions model a \texttt{try} block: a \texttt{CATCH} instruction to represent the start of the \texttt{try} block, the instructions representing the code inside the \texttt{try} block, a \texttt{CATCH} instruction to represent the end of the \texttt{try} block and a GOTO instruction targeting the instructions after the \texttt{try} block. Each catch statement is represented using a label, the instructions representing the exception handling and a GOTO instruction targeting the instructions after the \texttt{catch} block.

We use the same \texttt{CATCH} instruction to mark the beginning and end of the \texttt{try} block. However, \texttt{CATCH} instructions at the beginning and at the end differ by the information they hold; the \texttt{CATCH} instruction that marks the beginning of a \texttt{try} block has a map from the types of the catch statements and their labels in the GOTO program, while the second \texttt{CATCH} instruction has an empty map. The GOTO instruction targeting the instructions after the \texttt{catch} block shall be called in case no exception is thrown. The GOTO instructions at the end of each \texttt{catch} are called so that only the instructions of the current \texttt{catch} is executed, as shown in Fig.~\ref{figure:try-catch-goto}.
   \begin{figure}[h!]
   \centering
    \subfloat[{Try-catch example of throwing an integer exception.}\label{figure:try-catch}]{%
      \input{try-catch-example.tex}
    }
    \hspace*{1cm}%
    \subfloat[{GOTO instructions.}\label{figure:try-catch-goto-instructions}]{%
      \input{try-catch-goto.tex}
    }
    \caption{Example of try-catch conversion to GOTO instructions.}
    \label{figure:try-catch-goto}
  \end{figure}

During the SSA generation, when the first \texttt{CATCH} instruction is found, the map is stacked because there might be nested \texttt{try} blocks. If an exception is thrown, ESBMC encodes the jump to a catch statement according to the rules defined in Table~\ref{table:rules-catch}, including a jump to an invalid \texttt{catch} that triggers a verification error, i.e., it represents an exception thrown that can not be caught. If a suitable exception handler is found, then the thrown value is assigned to the \texttt{catch} variable (if any); otherwise, if there exists no valid exception, an error is reported. If the second \texttt{CATCH} instruction is reached and no exception was thrown, the map is freed for memory efficiency. The \texttt{try} block is handled as any other block in a C++ program. Destructors of variables in the stack are called by the end of the scope. Furthermore, by encoding throws as jumps, we also correctly encode memory leaks. For example, suppose an object is allocated inside a \texttt{try} block, and an exception is thrown and handled. In that case, it will leak unless the reference to the allocated memory is somehow tracked and freed.

Our symbolic engine also keeps track of \textit{function frames}, i.e., several pieces of information about the function it is currently evaluating, including arguments, recursion depth, local variables, and others. These pieces of information are essential not only because we want to handle recursion or find memory leaks but also allow us to connect exceptions thrown outside the scope of a function and handle exception specification (as described in Section~\ref{sec:exception-specification}).

%%%%%%%%%%%%%%%%%%%%%%%
\subsection{Exception Specification}
\label{sec:exception-specification}
%%%%%%%%%%%%%%%%%%%%%%%

The exception specification (illustrated in Fig.~\ref{figure:exception-especification-example}) defines which exceptions can be thrown by a function or method (including constructors). It is formed by an exception list and can be empty, i.e., the function or method cannot throw an exception. Exceptions thrown and handled inside a function or method are not affected by the exception specification.
%Fig.~\ref{figure:exception-especification-example}
%shows an example of the exception specification usage.
%
\begin{wrapfigure}[14]{r}{9cm}
\centering
\vspace{10pt}
\centering
\begin{minipage}{0.53\textwidth}
\begin{lstlisting}
/* function 1 can throw exceptions
   of type int and float */
void func1() throw(int, float) {
  ...
}
/* function 2 can't throw an exception */
void func2() throw() {
  try {
    /* OK, exception handled inside func2's scope */
    throw 1;
  }
  catch(int) {
    /* error handling for integer exceptions */
  }
}
\end{lstlisting}
\end{minipage}
\caption{Example of exception specification.}
\label{figure:exception-especification-example}
\end{wrapfigure}

To support the verification of programs with exception specifications, an instruction  \texttt{THROW\_DECL} is inserted at the beginning of the given function or method. This instruction contains a list of allowed exceptions that are checked whenever an exception is thrown outside the scope of the function or method. Similar to the  \texttt{catch} map, they are stacked due to the possibility of nested exception specifications and are freed at the end of the function or method.

An exception thrown from inside a function follows the same rules defined in Table~\ref{table:rules-catch}. Exception specifications check any exception thrown outside the function scope. If the type of the exception was not declared in the exception specialization, a different exception is raised and a separate path in the program is taken: the unexpected handler.\\

%---------------------------------------------------------------------------
\subsection{Terminate and Unexpected Handlers}
\label{sec:terminate-and-unexpected}
%---------------------------------------------------------------------------

During the exception handling process, errors can occur, causing the process to be aborted for any given reason (\textup{e.g.}, throwing an exception outside a \texttt{try} block or not catching a thrown exception). When this happens, the terminate handler is called.
   \begin{figure}[h!]
   \centering
    \subfloat[{Terminate functions: A) function type definition; B) Default terminate behavior; C) Set the default behavior; D) Function $set\_terminate$; E) Model for function terminate.}\label{figure:terminate-model}]{%
      \input{terminate.tex}
    }
    \hspace*{1cm}%
    \subfloat[{Unexpected functions: A) function type definition; B) Default unexpected  behavior; C) Set the default behavior; D) Function $set\_unexpected$; E) Model for function unexpected.}\label{figure:unexpected-model}]{%
      \input{unexpected.tex}
    }
    \caption{Examples of terminate and unexpected handlers.}
    \label{figure:terminate-unexpected}
  \end{figure}

Fig.~\ref{figure:terminate-model} shows the terminate handler implementation. The terminate handler is a function that has the default behavior of calling the {\tt abort} function. However, this behavior can be slightly changed by the developer, using the function {\tt set\_terminate(f)}, where $f$ is a function pointer to a function that has no parameter and no return value (type {\tt void}). By setting the new terminate function, it will be called before the {\tt abort} function.

For the verification of programs that override the terminate handler, we define a  function {\tt \_\_default\_terminate()}, as illustrated in Fig.~\ref{figure:terminate-model}, that contains the default termination behavior, calling {\tt abort}. ESBMC also keeps a global function pointer to the terminate function, which can either point to the default behavior or the user-defined behavior. Finally, when the terminate function is called, we should guarantee that the {\tt abort} function will be called, even if the terminate function is replaced (as shown in label $E$ in Fig.~\ref{figure:terminate-model}).

However, there is one case where the unexpected handler is called instead of the terminate handler. When an exception not allowed by the exception specification (Section~\ref{sec:exception-specification}) is thrown by a function or method, when this happens, the unexpected handler is called.

The unexpected handler works similarly to the terminate handler. It will either call terminate or re-throw the not allowed exception. Similar to {\tt set\_terminate}, there exists a function {\tt set\_unexpected(f)}, where {\tt f} is function pointer to a function that has no parameter and no return value (type {\tt void}).

Fig.~\ref{figure:unexpected-model} illustrates the unexpected handler implementation. The default behavior is to re-throw the thrown exception, and, as the exception specification already forbids it, we should call terminate to finish the program. ESBMC also keeps a global function pointer to the unexpected function, which either points to the default behavior or the user-defined behavior. If the unexpected handler was replaced, we must still guarantee that an exception will be thrown, so the forbidden exception will be re-thrown (as shown in line $27$ in Fig.~\ref{figure:unexpected-model}). If the replaced unexpected function throws an exception that is not forbidden by the function, the code will not terminate.
\begin{wrapfigure}[17]{l}{8cm}
\centering
%\vspace{-12pt}
\begin{minipage}{0.45\textwidth}
\begin{lstlisting}
#include <exception>
#include <cassert>
using namespace std;

void myunexpected () {
  throw;
}

void myfunction () throw (int, bad_exception) {
  throw 'x';
}

int main (void) {
  set_unexpected (myunexpected);
  try {
    myfunction();
  }
  catch (int) { assert(0); }
  catch (bad_exception be) { return 1; }
  return 0;
}
\end{lstlisting}
\end{minipage}
\caption{Fragment of code using bad exception.}
\label{figure:unexpected-badexception}
\end{wrapfigure}

Finally, we also need to model the unexpected behavior when using {\tt bad\_exception}. Fig.~\ref{figure:unexpected-badexception} shows an example of code using {\tt bad\_exception}. In this example, the user replaced the unexpected function with a function containing a re-throw. The code then calls {\tt myfunction()}, which tries to throw a forbidden {\tt char} exception. At this moment, {\tt myunexpected} function is called and tries to re-throw the {\tt char} exception, which is forbidden. ESBMC matches the compiler's behavior and checks whether {\tt bad\_exception} is one of the allowed exceptions in the exception specification; if this is true, a {\tt bad\_exception} exception will be thrown instead of the original forbidden exception.

%---------------------------------------------------------------
\section{Experimental Evaluation}
\label{sec:experimental-evaluation}
%---------------------------------------------------------------

Our experimental evaluation compares ESBMC against LLBMC and DIVINE regarding correctness and performance in the verification process of C++03 programs; DIVINE was developed by Baranov\' {a}~\textup{et al.}
~\cite{divine4}, and LLBMC was developed by Merz, Falke, and Sinz~\cite{MerzFS12}. Section~\ref{subsec:data} shows a detailed
description of all tools, scripts, and benchmark dataset,
while Section~\ref{subsec:discussion} presents the results and our evaluation.
Our experiments are based on a set of publicly available benchmarks. All tools, scripts,
benchmarks, and results of our evaluation are
available on a replication package~\cite{cppzenodo}, including {\bf all data to generate the percentages}.
More information about ESBMC is also available at the project's webpage \url{http://esbmc.org/}.

%---------------------------------------------------------------
\subsection{Experimental Design, Materials and Methods}
\label{subsec:data}
%---------------------------------------------------------------

Our experiments aim at answering two experimental questions regarding {\it correctness} and {\it performance} of ESBMC:
\begin{itemize}
 \item[\it i.] (EQ-I) How accurate is ESBMC when verifying the chosen C++03 programs?
 \item[\it ii.] (EQ-II) How does ESBMC performance compare to other existing model checkers?
\end{itemize}
To answer both questions, we evaluate all benchmarks with ESBMC $v2.1$, DIVINE $v4.3$, and LLBMC $v2013.1$.
ESBMC $v2.1$ contains the last stable version of our C++ front-end, since the changes necessary to introduce a new C front-end on ESBMC $v3.0$ were disruptive. The new C front-end is based on the clang's AST~\cite{esbmc2018}, which completely changes the way ESBMC processes source files. Update the C++ front-end to also use clang’s AST is part of our future work (cf. Section~\ref{sec:conclusion}).
We also applied CBMC~\cite{Clarke04} ($v5.3$) in our benchmark set. However, we do not detail the results in the experimental evaluation because the tool aborts during parser in 1,500 cases and reproduces false-negative results in the remaining 3. The vast majority of our benchmarks use STL functionalities, which CBMC does not support. The lack of support for C++ features in CBMC was also reported by Merz \textup{et al.}~\cite{MerzFS12}, Monteiro \textup{et al.}~\cite{STVR2017}, and Ramalho \textup{et al.}~\cite{Ramalho13}.

To tackle modern aspects of the C++ language, the comparison is based on a benchmark dataset that consists of 1,513 C++03 programs. In particular, 290 programs were extracted from the book ``C++ How to Program''~\cite{Deitel}, 432 were extracted from C++ Resources Network~\cite{CppReference12}, 16 were extracted from NEC Corporation~\cite{NeclabsBenchmarkExceptions}, 16 programs were obtained from LLBMC~\cite{MerzFS12}, 39 programs were obtained from CBMC~\cite{Clarke04}, 55 programs were obtained from the GCC test suite~\cite{GCC}, and the others were developed to check several features of the C++ programming language~\cite{Ramalho13}. The benchmarks are split into 18 test suites: {\it algorithm} contains 144 benchmarks to check the Algorithm library functionalities; {\it cpp} contains 357 general benchmarks, which involves C++03 libraries for general use, such as I/O streams and templates; this category also contains the LLBMC benchmarks and most NEC benchmarks. The test suites {\it deque} (43), {\it list} (72), {\it queue} (14), {\it stack} (14), {\it priority\_queue} (15), {\it stream} (66), {\it string} (233), {\it vector} (146), {\it map} (47), {\it multimap} (45), {\it set} (48), and {\it multiset} (43) contain benchmarks related to the standard template containers. The category {\it try\_catch} contains 81 benchmarks to the exception handling and the category {\it inheritance} contains 51 benchmarks to check inheritance and polymorphism mechanisms. Finally, the test suites {\it cbmc} (39), {\it templates} (23) and {\it gcc-template} (32) contain benchmarks from the GCC\footnote{{\url{https://github.com/nds32/gcc/tree/master/gcc/testsuite/}}} and CBMC\footnote{{\url{https://github.com/diffblue/cbmc/tree/develop/regression}}} test suite, which are specific to templates.

Each benchmark is tested and manually inspected in order to identify and label bugs. Thus, $543$ out of the $1,513$ benchmarks contain bugs ({\it i.e.}, $35.89$\%) and 970 are bug-free ({\it i.e.}, $64.11$\%). This inspection is essential to compare verification results from each model checker and properly evaluates whether real errors were found. We evaluate three types of properties: (i) memory-safety violations (e.g., arithmetic overflow, null-pointer dereferences, and array out-of-bounds), (ii) user-specified assertions, and (iii) proper use of C++ features (e.g., exception-handle violations). We only exclude LLBMC from the evaluation of exception handling since the tool does not support this feature. All tools support all the remaining features and properties under evaluation.

All experiments were conducted on a computer with an i7-4790 processor, 3.60GHz clock, with 16GB RAM and Ubuntu 14.04 64-bit OS. ESBMC, LLBMC, and DIVINE were set to a time limit of 900 seconds (i.e., 15 minutes) and a memory limit of 14GB. All presented execution times are CPU times, i.e., only the elapsed periods spent in the allocated CPUs. Furthermore, memory consumption is the amount of memory that belongs to the verification process and is currently present in RAM (i.e., not swapped or otherwise not-resident). Both CPU time and memory consumption were measured with the ${\tt times}$ system call (POSIX system). Neither swapping nor turbo boost was enabled during experiments and all executed tools were restricted to a single process.

The tools were executed using three scripts: the first one for ESBMC,\footnote{esbmc *.cpp --unwind $B$ --no-unwinding-assertions -I ~/libraries/} which reads its parameters from a file and executes the tool; the second one for LLBMC, which first compiles the program to bitcode, using clang,\footnote{clang++ -c -g -emit-llvm *.cpp -fno-exceptions -o main.bc}~\cite{CLANG} then it reads the parameters from a file and executes the tool;\footnote{llbmc *.o -o main.bc --ignore-missing-function-bodies --max-loop-iterations=$B$ --no-max-loop-iterations-checks} and the last one for DIVINE, which also first pre-compiles the C++ program to bitcode, then performs the verification on it.\footnote{divine verify *.cpp} The loop unrolling defined for ESBMC and LLBMC ({\it i.e.}, the $B$ value) depends on each benchmark. In order to achieve a fair comparison with ESBMC, an option from LLBMC had to be disabled. LLBMC does not support exception handling and all bitcodes were generated without exceptions (i.e., with the ${\tt -fno-exceptions}$ flag of the compiler). If exception handling is enabled, then LLBMC always aborts the verification process.

%%%%%%%%%%%%%%%%%%%%%%%%
\subsection{Results \& Discussion}
\label{subsec:discussion}
%%%%%%%%%%%%%%%%%%%%%%%%

\begin{figure*}[b] \centering
\includegraphics[width=\textwidth]{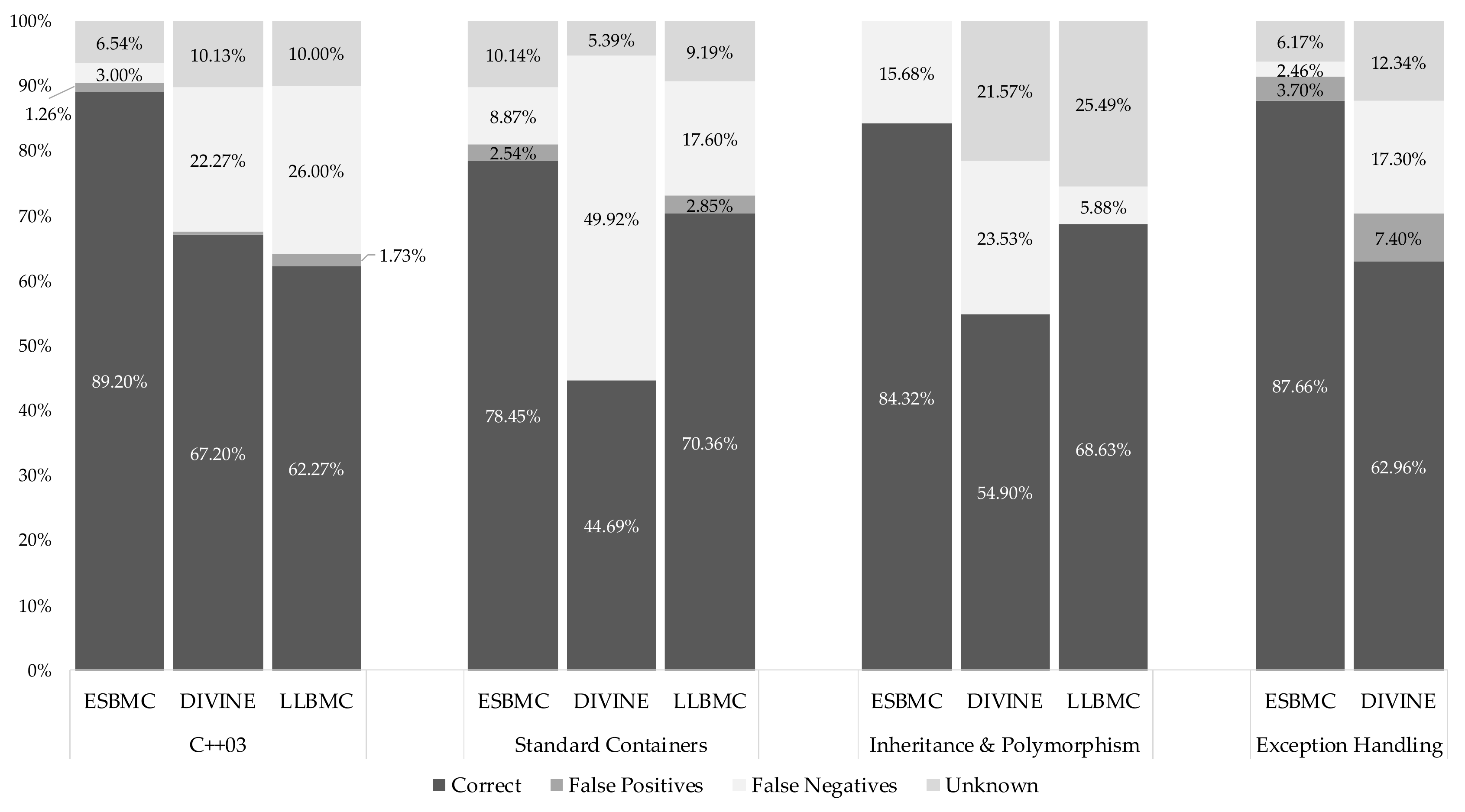}
\caption{{Experimental results for ESBMC $v2.1$, DIVINE $v4.3$, and LLBMC $v2013.1$.}}
\label{figure:tools-results}
\end{figure*}

In this section, we present the results using percentages
(concerning the 1,513 C++ benchmarks), as shown in Fig.~\ref{figure:tools-results}.
{\it Correct} represents the positive results, i.e., percentage of benchmarks with and without bugs correctly verified.
{\it False positives} represent the percentage of benchmarks reported as correct, but they are incorrect; similarly,
{\it False negatives} represent the percentage of benchmarks reported as incorrect, but that are correct. Finally, {\it Unknown} represents the
benchmarks where each tool aborted the verification process due to internal errors, timeout (\textup{i.e.}, the tool
was killed after $900$ seconds) or a memory out (\textup{i.e.}, exhausted the maximum memory allowed of $14$GB).
In the Exception Handling category, LLBMC is excluded since it does not support this feature;
if exception handling is enabled, then LLBMC continuously aborts the verification process.
%Appendix~\ref{appendix:experiments} gives the data to generate the percentages.
Furthermore, to better present the results
of our experimental evaluation, the test suites were grouped into four categories:

\begin{itemize}
\item {Standard Containers -- formed by {\it algorithm}, {\it deque}, {\it vector}, {\it list}, {\it queue}, {\it priority\_queue}, {\it stack}, {\it map}, {\it multimap}, {\it set} and {\it multiset} test suites (631 benchmarks);}
\item {Inheritance \& Polymorphism -- formed by the {\it inheritance} test suite (51 benchmarks).}
\item {Exception Handling -- formed by the {\it try\_catch} test suite (81 benchmarks);}
\item {C++03 -- formed by {\it cpp}, {\it string}, {\it stream}, {\it cbmc}, {\it gcc-templates} and {\it templates} test suites (750 benchmarks).}
\end{itemize}

On the Standard Containers category (see Fig.~\ref{figure:tools-results}), ESBMC presented the best results and reached a successful verification rate of $78.45$\%, while LLBMC reported $70.36$\% and DIVINE $44.69$\%. ESBMC's noticeable results for containers are directly related to its COM. The majority of the benchmarks for this category contain standard assertions to map the support of container-based operations, e.g., to check whether the ${\tt operator[]}$ from a ${\tt vector}$ object is called with an argument out of range, which is undefined behavior~\cite{ISO14882:2003}. We place standard C++ assertions in the benchmarks to evaluate how each verifier handles container-based operations. ESBMC reports a false-positive rate of $2.54$\% and a false-negative rate of $8.87$\%, which is due to internal implementation issues during pointer encoding ({\it cf.}, Section~\ref{sec:cpp-operational-model}). We are currently working to address them in future versions. ESBMC also reported $10.14$\% of unknown results due to limitations in templates-related features such as SFINAE~\cite{ISO14882:2003} and nested templates. LLBMC reports a false-positive rate of $2.85$\% and a false-negative rate of $17.60$\%, mostly related to erroneously evaluating assertions (\textup{e.g.}, assertions to check whether a container is empty or it has a particular size). It also reports an unknown rate of $9.19$\% regarding timeouts, memory outs, and crashes when performing formula transformation~\cite{MerzFS12}. DIVINE does not report any timeout, memory out, or false-positive results for this category, but an expressive false-negative rate of $49.92$\%, resulting from errors to check assertions (similarly to LLBMC). DIVINE also reports an unknown rate of $5.39\%$ due to errors with pointer handling, probably due to imprecise (internal) encoding.

On the Inheritance \& Polymorphism category (see Fig.~\ref{figure:tools-results}), ESBMC presented the best results and reached a successful verification rate of $84.32$\% while LLBMC reported $68.63$\% and DIVINE $54.90$\%. ESBMC does not report any timeout or memory out, but it reports a false-negative rate of $15.68$\%, due to implementation issues to handle pointer encoding. LLBMC does not report any false positives, timeouts, or memory outs results. However, it reports a false-negative rate of $5.88$\%, which is related to failed assertions representing functional aspects of inherited classes. It also reported an unknown rate of $25.49$\% regarding multiple inheritance. DIVINE does not report any timeout, memory out, or false-positive results for this category, but a false-negative rate of $23.53\%$ and an unknown rate of $21.57\%$, which is a result of errors when handling dynamic casting, virtual inheritance, multiple inheritance, and even basic cases of inheritance and polymorphism.

On the Exception Handling category (see Fig.~\ref{figure:tools-results}), ESBMC presented the best results and reached a successful verification rate of $87.66$\% while DIVINE reported $62.96$\%. ESBMC does not report any timeout or memory out, but it reports a false-positive rate of $3.70$\% and a false-negative rate of $2.47$\%. These bugs are related to the implementation of rule $r_6$ from Table~\ref{table:rules-catch} in ESBMC, i.e., ``catches an exception if the type of the thrown exception $e$ can be converted to the type of the catch $h$, either by qualification or standard pointer conversion''; we are currently working on fixing these issues. ESBMC also presents an unknown rate of $3.70$\% due to previously mentioned template limitations. DIVINE does not report any timeout or memory out. However, it reports a false-positive rate of $7.40$\% and a false-negative rate of $17.30$\%. It incorrectly handles re-throws, exception specification, and the unexpected as well as terminate function handlers. DIVINE also presents an unknown rate of $12.34$\% due to errors when dealing with exceptions thrown by derived classes, instantiated as base classes, which is probably related to the imprecise encoding of \textit{vtables}.

To evaluate how these model checkers perform when applied to general C++03 benchmarks, we evaluate them against the category C++03. In this category, model checkers deal with benchmarks that make use of the features discussed in this paper (\textup{e.g.}, exception handling and containers), and a wider range of libraries from the STL, manipulation of strings and streams, among other C++03 features. ESBMC presented the highest successful verification rate, $89.20$\%, followed by DIVINE $67.20$\% and LLBMC $62.27$\%. The successful expressive rate of ESBMC in this category not only correlates to its support for core C++03 features (\textup{i.e.}, templates, inheritance, polymorphism, and exception handling) or its ability to check functional aspects of the standard containers but also because COM contains abstractions for all standard libraries shown in Table~\ref{table:overview-com}. For instance, the operational model for the string library enables ESBMC to achieve a success rate of $99.14$\% in the {\it string} test suite, which contains benchmarks that target all methods provided in C++03 for ${\tt string}$ objects. Note that running ESBMC without COM over the benchmarks, $98.08$\% fail since the majority uses at least one standard template library. ESBMC does not report any memory out, but it reports a false-positive rate of $1.26$\%, a false-negative rate of $3.00$\%, and an unknown rate of $6.54$\%, which are all due to the same issues pointed by the previous experiments.
DIVINE does not report any false positives, timeout, or a memory out, but a false-negative rate of $22.27\%$, which is a result of errors when checking assertions representing functional properties of objects across all STL (similar to LLBMC). DIVINE reports one false positive regarding the instantiation of function template specialization and an unknown rate of $10.13$\% due to crashes when handling pointers.
LLBMC reports a false-positive rate of $1.73$\% and a false-negative rate of $26.00$\%, which is related to errors when checking assertions that represent functional properties of objects (\textup{e.g.}, asserting the size of a ${\tt string}$ object after an operation) or dealing with ${\tt stream}$ objects in general. It also reported an unknown rate of $10.00$\%, mainly regarding operator overloading errors and the ones mentioned in the previous categories.

A small number of counterexamples generated by the three tools were manually checked, but we understand that this is far from ideal. The best approach is to use an automated method to validate the counterexample, such as the witness format proposed by Beyer et al.~\cite{svcomp2015}; however, the available witness checkers do not support the validation of C++ programs. Implementing such a witness checker for C++ would represent a significant development effort, which we leave it for future work.
\begin{figure}[ht] \centering
\includegraphics[scale=0.47]{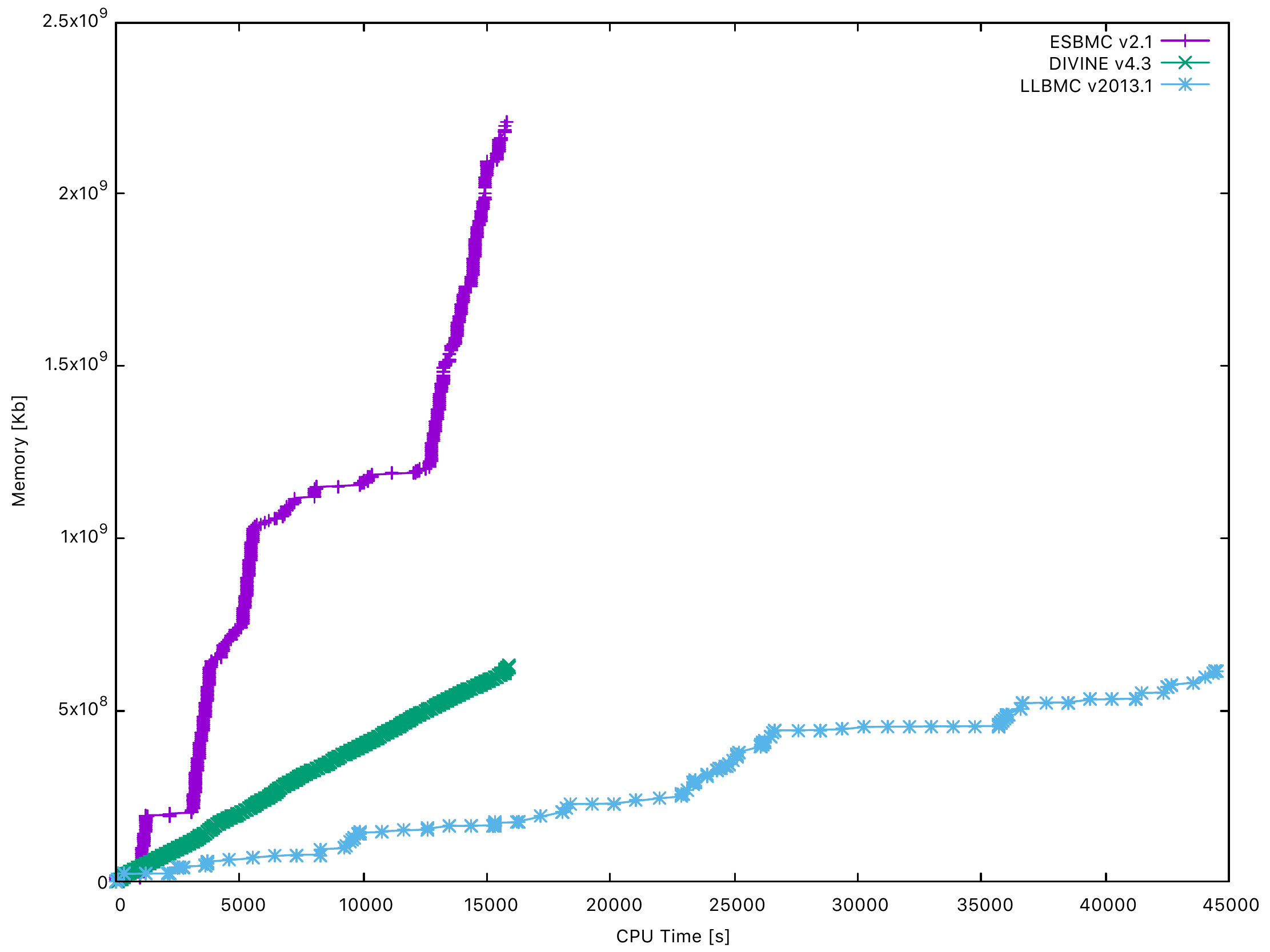}
\caption{Comparison of accumulative verification time and accumulative memory consumption among ESBMC $v2.1$, DIVINE $v4.3$, and LLBMC $v2013.1$ throughout the verification process of all benchmarks.}
\label{figure:tools-results-time}
\end{figure}

Fig.~\ref{figure:tools-results-time} illustrates the accumulative verification time and memory consumption for the tools under evaluation. All the tools take more time to verify the test suites \textit{algorithm}, \textit{string}, and \textit{cpp}, due to a large number of test cases and the presence of pointers and iterators. ESBMC is the fastest of the three tools, $3.2$ times faster than LLBMC and only $155.7$ seconds faster than DIVINE. In terms of verification time, DIVINE is the only tool that did not use more than the defined limit of $900$ seconds, while ESBMC and LLBMC aborted due to
timeout in $4$ and $25$ benchmarks, respectively. DIVINE is the only tool that did not use more than the defined limit of $14$GB per benchmark in terms of memory consumption. At the same time, ESBMC and LLBMC aborted due to exhaustion of the memory resources in $3$ and $11$ of them, respectively. Even so, LLBMC consumes less memory overall ($614.92$GB) when compared to DIVINE ($627.97$GB) and ESBMC ($2,210.91$GB).

Overall, ESBMC achieved the highest success rate of $84.27$\% in $15,761.90$ seconds (approximately $4$ hours and 23 minutes), faster than LLBMC and DIVINE, which positively answers our experimental questions EQ-I and EQ-II. LLBMC correctly verified $62.52$\% in $50,564.10$ seconds (approximately $14$ hours) and can only verify the programs if exception handling is disabled, which is not a problem for both ESBMC and DIVINE. DIVINE correctly verified $57.17$\% in $15,917.60$ seconds (approximately $4$ hours and $26$ minutes). Regarding memory usage, ESBMC has the highest usage among the three tools, which is approximately $3.5$ times higher than DIVINE and LLBMC, respectively. This high consumption is due to the generation process of SSA forms (\textup{cf.}, Section~\ref{sec:type-checking}). However, its optimization is under development for future versions.

In conclusion, our experimental evaluation indicates that ESBMC outperforms two state-of-the-art model checkers, DIVINE and LLBMC, regarding the verification of inheritance, polymorphism, exception handling, and standard containers. The support for templates in ESBMC needs improvements. However, the current work-in-progress clang front-end will not only cover this gap (because clang will instantiate all the templates in the program) but will also allow ESBMC to handle new versions of the language (\textup{e.g.}, C++11). Even with its current support for templates, our experimental results allow us to conclude that ESBMC represents the state-of-the-art regarding applying model checking in C++ programs.

%-----------------------------------------------------
\subsection{Sniffer Application}
\label{sec:sniffer-application}
%-----------------------------------------------------

This section describes the results of the verification process using ESBMC and LLBMC in a \textit{sniffer} program. We were unable to use DIVINE to verify the code because the tool does not offer support for the verification of some libraries used in the program (\textup{e.g.}, {\it boost}~\cite{boost}), which makes the verification process an infeasible task{, i.e., DIVINE would report incorrect results}. Nokia Institute of Technology (INdT) made this program available. Sniffer is responsible for capturing and monitoring traffic conditions of a network, which supports Message Transfer Part Level $3$ User Adaptation Layer (M3UA)~\cite{M3UA}. This service offers the transport of SS7 protocols (Signalling System $No. 7$) and makes use of the services provided by the Stream Control Transmission Protocol (SCTP). The \textit{Sniffer} program contains $20$ classes, $85$ methods, and $2,800$ lines of C++ code.
\begin{wrapfigure}[9]{r}{8cm}
\centering
%\vspace{-12pt}
\begin{minipage}{0.45\textwidth}
\begin{lstlisting}
int PacketM3UA::getPayloadSize() {
  return ntohs(m3uaParamHeader->paramSize)
          - (M3UA_PROTOCOL_DATA_HEADER_SIZE
          + M3UA_PARAMETER_HEADER_SIZE);
}
\end{lstlisting}
\end{minipage}
\caption{Arithmetic overflow in the ${\tt typecast}$ operation from
the ${\tt getPayloadSize}$ method.}
\label{figure:PacketM3UA}
\end{wrapfigure}

The following properties were verified in the \textit{sniffer} program:
arithmetic under- and overflow, division by zero, and
array bounds violation. Due to confidentiality issues, we were only able to
verify $50$ of $85$ methods since INdT did not provide some classes required by
the unverified methods. From the verified code, ESBMC was able to identify five errors, related to arithmetic under- and overflow while
LLBMC was able to identify only three of them. All errors were reported to
developers, who confirmed them.
As an example of an error found, Fig.~\ref{figure:PacketM3UA} shows the
${\tt getPayloadSize}$ method from the ${\tt PacketM3UA}$ class. In this
method, an arithmetic overflow can occur. The method returns
${\tt ntohs}$, an ${\tt unsigned \: int}$, but the ${\tt getPayloadSize}$ method
must return a ${\tt signed \: int}$. In this case, a possible solution is to
change the return type of the ${\tt getPayloadSize}$ method to ${\tt unsigned \: int}$.

%----------------------------------
\section{Related Work}
\label{sec:related-work}
%----------------------------------

Conversion of C++ programs into another language makes the verification process easier since C++ model checkers are still in the early development stages. There are more stable verification tools written for other programming languages, such as C~\cite{svcomp2019}. This conversion, however, can unintentionally introduce or hide errors in the original program. In particular, the converted program's verification may present different results if compared to the verification of the original C++ program, unless we check the equivalence of both the original and the modified program~\cite{GodlinS13}, {which can become undecidable in the presence of unbounded memory usage}.

\begin{table}[ht]
\renewcommand\arraystretch{1.15}
\setlength{\tabcolsep}{4pt}
\caption{Related work comparison.}
\begin{center} {\scriptsize
\begin{tabular}{| >{\centering\arraybackslash}m{2.7cm} | >{\centering}m{1.5cm} |c| >{\centering}m{1.9cm} | >{\centering}m{1.9cm} | c |}
\hline
Related work & Conversion to intermediate languages & \multicolumn{4}{c|} {C++ Programming Language Support} \\
\cline{3-6}
   &    & Templates & Standard Template Libraries & Inheritance \& Polymorphism &  Exception Handling  \\
\hline
\hline
LLBMC~\cite{llbmc2013} & LLVM & Yes & Yes & Yes & No \\ \hline                            % LLBMC
SATABS~\cite{Blanc07} & No & Yes & Yes & No & No \\ \hline                                    % SATABS
%Prabhu~\textup{et al.}~\cite{Prabhu11} & ANSI-C & Yes & Not mentioned & Yes & Yes \\ \hline    % CIL
CBMC~\cite{Clarke04} & No & Yes & No & No & No \\ \hline                                   % CBMC
DIVINE~\cite{divine4} & LLVM & Yes & Yes & Yes & Yes \\ \hline                     % DIVINE
%Cppcheck~\cite{CPPChecker} & No & Yes & Yes & No & Yes \\ \hline    % CppCheck
{\bf ESBMC v2.0}~\cite{Ramalho13} & {\bf No} & {\bf Yes} & {\bf Yes} & {\bf Yes} & {\bf Yes} \\ \hline
\end{tabular} }
\end{center}
\label{tab:comparison}
\end{table}

When it comes to the verification of C++ programs, most of the model checkers
available {in the literature} focus {their verification approach}
on specific {C++} features, such as exception handling, and end up
neglecting other features of equal importance, such as the verification of the
STL~\cite{divine2016, divine2017}.
Table~\ref{tab:comparison} shows a comparison among other studies available in
the literature and our approach.

Merz, Falke, and Sinz~\cite{MerzFS12, llbmc2013} describe LLBMC, a tool that uses BMC to verify C++ programs. The tool first converts the program into LLVM intermediate representation, using clang~\cite{CLANG} {as an off-the-shelf front-end}. This conversion removes high-level information about the structure of C++ programs (\textup{e.g.}, the relationship between classes). However, the code fragments that use the STL are inlined, which simplifies the verification process. From the LLVM intermediate representation, LLBMC generates a {quantifier-free logical formula based on bit-vectors}. This formula is further simplified and passed to an SMT solver for verification. The tool does not verify programs with exception handling, making it challenging to verify C++ programs realistically since exceptions must be disabled during the generation of the LLVM intermediate representation. The biggest difference between the tool described by the authors and the purpose of this work is related to the beginning of the verification process. In LLBMC, the conversion of the source program into an intermediate representation LLVM is required. The biggest obstacle to this approach is the need for a constant tool adjustment to new versions of the LLVM intermediate representation that the clang generates. For instance, a symbolic virtual machine built on top of the LLVM compiler, named as KLEE~\cite{Cadar:2008:KUA:1855741.1855756}, still uses an old version of LLVM (v3.4) due to the significant effort to update its internal structure.

Blanc, Groce, and Kroening~\cite{Blanc07} describe the verification of C++ programs using containers via predicate abstraction. A simplified operational model using Hoare logic is proposed to support C++ programs that make use of the STL. The purpose of the operational model is to simplify the verification process using the SATABS tool~\cite{satabs}. SATABS is a verification tool for C and C++ programs that supports classes, operator overloading, references, and templates (but without supporting partial specification). In order to verify the correctness of a program, the authors show that it is sufficient to use an operational model by proving that, if the pre- and postconditions hold, the implementation model also holds. The approach is efficient in finding trivial errors in C++ programs. The preconditions are modeled to verify the library containers using an operational model similar to the ESBMC tool's model for the same purpose. Regarding the operational model, the authors present only preconditions. In contrast, our operational model verifies preconditions and replicates the STL behavior, which increases the range of applications that can be adequately verified by the tool (\textup{i.e.}, postconditions).

Clarke, Kroening, and Lerda~\cite{Clarke04} present CBMC, which implements BMC for  C/C++ programs using SAT/SMT solvers. CBMC uses its parser, based on Flex/Bison~\cite{CordeiroFM12}, to build an AST. The type-checker of CBMC's front-end annotates this AST with types and generates a language-independent intermediate representation of the original source code. The intermediate representation is then converted into an equivalent GOTO-program (\textup{i.e.}, control-flow graphs) that the symbolic execution engine will process. ESBMC improves the front-end, the GOTO conversion and the symbolic execution engine to handle the C++$03$ standard. CBMC and ESBMC use two functions $\cal C$ and $\cal P$ that compute the $constraints$ (\textup{i.e.}, assumptions and variable assignments) and $properties$ (\textup{i.e.}, safety conditions and user-defined assertions), respectively. Both tools automatically generate safety conditions that check for arithmetic overflow and underflow, array bounds violations, and null pointer dereferences, in the spirit of Sites' clean termination~\cite{Sites74}. Both functions accumulate the control-flow predicates to each program point and use these predicates to guard both the constraints and the properties so that they properly reflect the semantics of the program. A VC generator (VCG) then derives the verification conditions from them. CBMC is a well-known model checker for C programs, but its support for C++ is rather incomplete (\textup{cf.} Section~\ref{sec:experimental-evaluation}). In particular, CBMC has problems instantiating template correctly and lacks support for STL, exception specialization and terminate/unexpected functions.

Baranov\'{a}~\textup{et al.}~\cite{divine4} present DIVINE, an explicit-state model checker to verify single- and multi-threaded programs written in C/C++ (and other input formats, such as UPPAAL\footnote{http://www.uppaal.org} and DVE\footnote{http://divine.fi.muni.cz/index.html}). Another language supported by DIVINE is the LLVM intermediate representation; for this reason, the base of its verification process is the translation of C++ programs into that representation. Using clang~\cite{CLANG} as front-end, DIVINE translates C++ programs into the LLVM intermediate representation, {thereby}, applying its implementation of the C and C++ standard libraries in order to ensure a consistent translation. Nonetheless, this translation process might cause some irregularities to the verification process once it loses high-level information about the C++ program structure (\textup{i.e.}, the relationship between the classes). To tackle such issues in the verification process of exception handling structures, {\v{S}}till, Ro{\v{c}}kai and Barnat~\cite{divine2017} propose a new API for DIVINE to properly map and deal with exception handling in C++ programs, based on a study about the C++ and LLVM exception handling mechanisms~\cite{divine2016}. The authors also claim DIVINE as the first model checker that can verify exception handling in C++ programs, as opposed to what has been stated by Ramalho~\textup{et al.}~\cite{Ramalho13}. However, ESBMC $v1.23$ (\textup{i.e.}, the version used by Ramalho~\textup{et al.}~\cite{Ramalho13}) is able to correctly verify the example presented by Ro{\v{c}}kai, Barnat and Brim~\cite{divine2017}, generating and verifying $10$ VCs in less than one second. Our experimental evaluation shows that ESBMC outperforms DIVINE in handling exceptions as well as for the support of standard containers, inheritance, and polymorphism (\textup{cf.} Section~\ref{sec:experimental-evaluation}).

%-=-=-=-=-=-=-=-=-=-=-=-=-=-=-=-=-=
\section{Conclusions \& Future Work}
\label{sec:conclusion}
%-=-=-=-=-=-=-=-=-=-=-=-=-=-=-=-=-=

We have described a novel SMT-based BMC approach to verify C++ programs using ESBMC.
We started with an overview of ESBMC's type-checking engine, which includes our approach to support templates (similar to conventional compilers) that replace the instantiated templates before the encoding phase. We also describe our type-checking mechanism to handle single and multiple inheritance and polymorphism in C++ programs. We then present the significant contributions of this work: the C++ operational model and the support for exception handling. We describe an abstraction of the standard template libraries, which replaces it during the verification process. The purpose is twofold: reduce complexity while checking whether a given program uses the STL correctly.
Finally, we present novel approaches to handle critical features of exception handling in C++ (e.g., unexpected and termination function handlers).

To evaluate our approach, we extended our experimental evaluation by approximately $36$\% if compared to our prior work~\cite{Ramalho13}. ESBMC is able to verify correctly $84.27$\%, in approximately $4$ hours, outperforming two state-of-art verifiers, DIVINE and LLBMC (\textup{cf.}, Section~\ref{sec:experimental-evaluation}).
ESBMC and DIVINE were also able to verify programs with exceptions enabled, a missing feature of LLBMC that decreases the verification accuracy of real-world C++ programs. Besides, ESBMC was able to find undiscovered bugs in the \textit{Sniffer} code, a commercial application of medium-size used in the telecommunications domain. The developers later confirmed the respective bugs. LLBMC was able to discover a subset of the bugs discovered by ESBMC, while DIVINE was unable to verify the application due to a lack of support for the Boost C++ library~\cite{boost}.

%----------------------------------------------
%\subsection{Correctness}
%\label{sec:correctness}
%----------------------------------------------

Our verification method depends on the fact that COM correctly represents the original STL. Indeed, the correctness of such a model to trust in the verification results is a significant concern~\cite{STVR2017, android:2012, android:2015, monteiro:2015, gpu2016, GarciaMCF16, ESBMCGPU2018}.
The STL is specified by the ISO International Standard ISO/IEC 14882:2003(E) – Programming Language C++~\cite{ISO14882:2003}. Similar to conformance testing~\cite{CamaraGMS09,CamaraCGM11}, to certify the correlation between STL and COM, we rely on the translation of the specification into assertions, which represents the pre- and post-conditions of each method/function in the SCL.
Although COM is an entirely new implementation, it consists in (reliably) building a simplified model of the related STL, using the C/C++ programming language through the ESBMC intrinsic functions (\textup{e.g.}, ${\tt assert}$ and ${\tt assume}$) and the original specification, which thus tends to reduce the number of programming errors. Besides, Cordeiro {\it et al.}~\cite{CordeiroFM12,CordeiroF11,Cordeiro10} presented the soundness for such intrinsic functions already supported by ESBMC. Although proofs regarding the soundness of the entire operational model could be carried out, it represents a laborious task due to the (adopted) memory model~\cite{Mehta:2005}. Conformance testing concerning operational models would be a suitable approach~\cite{STVR2017,CamaraCGM11} and represents a promising approach for future research.

For future work, we intend to extend ESBMC coverage in order to verify C++$11$ programs. The new standard is a huge improvement over the C++$03$, which includes the replacement of exception specialization by a new keyword ${\tt noexcept}$, which works in the same fashion as an empty exception specialization. The standard also presents new sequential containers (${\tt array}$ and ${\tt forward\_list}$), new unordered associative containers (${\tt unordered\_set}$, ${\tt unordered\_multiset}$, ${\tt unordered\_map}$ and ${\tt unordered\_multimap}$), and new multithreaded libraries (\textup{e.g.}, ${\tt thread}$) in which our COM does not yet support. Finally, we will develop a conformance testing procedure to ensure that our COM conservatively approximates the STL semantics.

Furthermore, we intend to improve the general verification of C++ programs, including improved support for templates. Although the current support of templates was sufficient to verify real-world C++ applications ({\it cf.}, Section~\ref{sec:experimental-evaluation}) it is still work-in-progress. For instance, the handling of SFINAE~\cite{ISO14882:2003} in ESBMC is limited, and limitations on the support of nested templates, as shown in the experiments, directly affect the verification process. This limitation is because template instantiation is notoriously hard, especially if we consider recent standards. Although our front-end can handle many real-world C++ programs, maintaining the C++ front-end in ESBMC is a herculean task. For that reason, we decided to rewrite our front-end using clang~\cite{CLANG} to generate the program AST. Importantly, we do not intend to use the LLVM intermediate representation but the AST generated by clang. In particular, if we use clang to generate the AST, then it solves several problems:
{\it (i)} the AST generated by clang contains all the instantiated templates so
 we only need to convert the instantiated classes/functions and ignore the
 generic version;
{\it (ii)} supporting new features will be as easy as adding a new AST conversion
 node from the clang representation to ESBMC representation;
{\it (iii)} we do not need to maintain a full C++ front-end since ESBMC will contain all libraries from clang. Thus, we can focus on the main goal of ESBMC, the SMT encoding of C/C++ programs.

We already took the first step towards that direction and rewrote the C front-end~\cite{esbmc2018}, and the C++ front-end is currently under development.

\end{document}